\documentclass[10pt, conference, letterpaper]{IEEEtran}
\IEEEoverridecommandlockouts
\usepackage{bm}
\usepackage{amsmath}
\usepackage{graphicx}
\usepackage[linesnumbered,vlined,ruled]{algorithm2e}
\usepackage{amsfonts,amssymb}
\usepackage{cases}
\usepackage{bm}

\usepackage{amsthm}
\usepackage{cite}
\usepackage{xcolor}
\usepackage{mathrsfs}
\usepackage{multicol}
\usepackage{subfigure}
\usepackage{float}
\usepackage{url}

\newtheorem{defn}{\textbf{Definition}}
\newtheorem{remark}{\emph{Remark}}

\newtheorem{theorem}{\textbf{Theorem}}
\newtheorem{lemma}{\textbf{Lemma}}

\newtheorem{problem}{\textbf{Problem}}

\newcommand{\drafty}[1]{\textcolor{black}{#1}}
\newcommand{\hstan}[1]{\textcolor{black}{#1}}

\newcommand{\ttan}[1]{\textcolor{black}{#1}}

\newcommand{\htan}[1]{\textcolor{black}{#1}}
\newcommand{\hzhua}[1]{\textcolor{black}{#1}}
\newcommand{\hzhuaa}[1]{\textcolor{black}{#1}}
\newcommand{\grt}[1]{\textcolor{black}{#1}}
\begin{document}

\title{Dynamic  Virtual Machine Management via Approximate Markov Decision Process\thanks{This is the full version of the paper appeared in INFOCOM'16. \newline \indent Part of Z. Han's work was done when he was visiting at Jinan University. \newline \indent  Contact Haisheng Tan at thstan@jnu.edu.cn.\newline \indent This work is supported in part by NSFC Grants 61502201, 61401192, 61472252, China 973 project (2014CB340303), Hong Kong RGC CRF Grant C7036-15G, and NSF-Guangdong Grant 2014A030310172.} }
\author{
    \IEEEauthorblockN{Zhenhua Han\IEEEauthorrefmark{1}\IEEEauthorrefmark{2}\IEEEauthorrefmark{4}, Haisheng Tan\IEEEauthorrefmark{2}, Guihai Chen\IEEEauthorrefmark{3}, Rui Wang\IEEEauthorrefmark{1}, Yifan Chen\IEEEauthorrefmark{1}, Francis C.M. Lau\IEEEauthorrefmark{4}}
    \IEEEauthorblockA{\IEEEauthorrefmark{1}	South U. of Sci. \& Tech. of China ~~~~~\IEEEauthorrefmark{2}	Jinan University, Guangzhou
    }
    \IEEEauthorblockA{~~~\IEEEauthorrefmark{3}	Shanghai Jiao Tong University~~~~~~~~~\IEEEauthorrefmark{4}	The University of Hong Kong
    }

}
\maketitle

\begin{abstract}
Efficient virtual machine (VM) management can dramatically reduce energy
consumption in data centers. Existing VM management algorithms fall into
two categories based on whether the VMs' resource demands are assumed to
be static or dynamic. The former category fails to maximize the resource
utilization as they cannot adapt to the dynamic nature of VMs' resource
demands. Most approaches in the latter category are heuristical and lack
theoretical performance guarantees.
In this work, we formulate dynamic VM
management as a large-scale Markov Decision Process (MDP) problem and derive an optimal solution. Our
analysis of real-world data traces supports our choice of the
modeling approach. However, solving the large-scale MDP problem suffers
from the curse of dimensionality. Therefore, we further exploit the special structure of the problem and propose an
approximate MDP-based dynamic VM management method, called MadVM. We
prove the convergence of MadVM and analyze the bound of its approximation
error. Moreover, MadVM can be implemented in a distributed system, which
should suit the needs of real data centers. Extensive
simulations based on two real-world workload traces show that MadVM
achieves significant performance gains over two existing baseline
approaches in power consumption, resource shortage and the number of VM
migrations. Specifically, the more intensely the resource demands fluctuate, the more MadVM outperforms.
\end{abstract}

\section{Introduction}



Virtual machine (VM) is a widely-used technology for data center
management. 
By adopting virtualization-based solutions, servers could generate VMs
according to users' requests for storage space, computing resources
(CPU cores or CPU time) and network bandwidth. Since multiple
VMs could co-exist in a single server, virtualization could
improve the utilization of the underloaded servers, which leads to
reduced power consumption as fewer servers are used. Energy-efficient resource
management for virtualization-based data centers has thus become an attractive
research area.


In the literature, energy-efficient VM
management in data centers can be divided into two categories based on
whether the resource demands of the VMs are static or dynamic over
time. If resource demand is treated as constant, the management problem can
be formulated as a bin-packing
problem~\cite{li2014let,kuo2014optimal,nathuji2007virtualpower,popa2012faircloud,alicherry2013optimizing,guo2013shadow,cohen2013almost}.
Since in reality resource demands of VMs are essentially dynamic, static
approaches would have low resource utilization. When the resource
demands of VMs are dynamic, it may happen that
insufficient resources are provided to the VMs (called service-level agreement (SLA) violation or
resource shortage).
Thus, the allocation and migration policies of VMs should be made
adaptive to such situations by jointly considering energy consumption and
resource shortage over time.
Jin et al. \cite{jin2012efficient} modeled the resource demands of VMs as
a normal distribution and proposed a stochastic bin-packing algorithm.
They assumed that the statistics of the stochastic demands could be known,
which is hard in practice.
Some other works (e.g., \cite{Bobroff2007,shen2011cloudscale}) predicted
the resource demands of VMs with historical data, which aimed mainly to
avoid resource shortage, but have not considered the possibility of
saving energy with VM migrations.
Beloglazov et al. \cite{beloglazov2010energy} proposed an algorithm for
VM consolidation that jointly considered energy consumption, QoS and
migration costs. They pursued the optimization of energy consumption at a
time point rather than over a long period. Chen et al.
\cite{chen2014consolidating} predicted the resource demand pattern for
the initial VM placement, and migrated the VMs when the prediction turned
out to be wrong. It is obvious that the accuracy of the prediction will degrade
over time, and so will the resource utilization.

When considering dynamic resource demand, existing approaches of VM
management are all centralized
\cite{jin2012efficient,beloglazov2010energy,chen2014consolidating,Bobroff2007,beloglazov2013managing,shen2011cloudscale}.
The data center would be in a panic when the centralized controller
breaks down. Moreover, most of the dynamic approaches are
heuristics-based,
hence lack sufficient theoretical performance \grt{guarantee}. In this paper, we
regard the VM management as a stochastic optimization problem. Our
analysis of real-world data traces shows that by properly choosing the
time-slot duration, the first-order transition probability of the VMs'
resource demands is quasi-static for a long period and non-uniformly
distributed, and hence the Markov chain model would be a simple and effective
tool to capture the temporal correlations of the demands. Therefore, we
adopt the Markov chain model to study the VM management problem, for which
we jointly consider energy consumption and resource shortage. Our
contributions can be summarized as follows:
      \begin{itemize}
\item We formulate the dynamic VM management problem as an infinite
horizon large-scale Markov Decision Process (MDP) problem with an
objective to minimize the long-term average cost, the weighted sum of the
power consumption and resource shortage. By solving the equivalent
Bellman's equation, we propose an optimal VM management method with a
time complexity that is exponential of the number of VMs.
\item We propose an approximate MDP-based dynamic VM management
algorithm, called MadVM. Moreover, we prove the convergence of MadVM and analyze the bound on the
gap from the optimal solution. Moreover, our approach can be
implemented in a distributed system with very little global information,
and can therefore improve the robustness and scalability of data centers.
\item We conduct detailed simulations based on two real-world
workload traces. The experimental results show that MadVM can achieve
significant performance gains over other existing algorithms in power
consumption, resource shortage and the number of VM migrations, i.e., while maintaining the near-$0$ resource shortage, MadVM can reduce the power consumption by up to 23\% and 47\%  (averagely 19\% and 42\%) compared with CompVM~\cite{chen2014consolidating} and CloudScale~\cite{shen2011cloudscale}, respectively.
      \end{itemize}


The remainder of this paper is organized as follows.  In Section
\ref{section:system}, we define the system model and the dynamic VM
management problem. In Section \ref{section:centralized-method}, we
derive optimal and approximate algorithms for the problem. In Section
\ref{section:performance}, we analyze the temporal correlation of
resource demands in Google data centers to validate our choice of using the
Markov chain model. We evaluate the performance of our
approach using the real-world data traces by comparing with other existing
approaches. We conclude the paper in Section \ref{section:conclusion}.

\section{System Model and Problem Definitions \label{section:system}}
In this section, we define the system model, the power consumption model as
well as the VM migration model, and formulate our dynamic VM management
problem. 
      \subsection{System Architecture}
We consider a large-scale data center with a collection of physical servers
(also called Physical Machines, PMs), where VMs are generated to run in a
server according to the users' requests. Each VM is allocated to one PM,
whereas a PM can be allocated multiple VMs. \htan{We assume there is a
centralized manager to handle the allocation and migration of VMs.}

For convenience, we assume that there are sufficient memory
and network capacity for the VMs in each server. Only the scheduling of
computational power (CPU time) is considered as it is the main factor
determining the power consumption of the servers \cite{buyya2010energy}.
\footnote{\hstan{Note that through extending the dimensions of the state
space $\mathbb Q$, our solution can be extended to multiple
resources allocation, e.g., the CPU time, memory, and network
bandwidth.}} We assume the resource capacity of each server is identical,
which is denoted as $T_r$. Different VM allocations will lead to
different CPU utilizations and power consumptions. Thus, the centralized
manager should be carefully designed to schedule allocation of VMs for
power conservation. A slotted scheduling framework is adopted, where time
is divided into a sequence of time-slots of the same duration. At the
beginning of a time-slot, each VM determines and submits its demand for
CPU to the manager.

Let $V_m \triangleq \{m_1,\ldots,m_{|V_m|}\}$ and $V_s \triangleq
\{s_1,\ldots,s_{|V_s|}\}$ denote the sets of VMs and PMs, respectively. Set
$R_{l}(t)$ as the demand for CPU time of VM $m_l$ at time $t$. Let $Y_i(t)$ denote the location of the VM $m_i$
at time-slot $t$, i.e. $Y_i(t) = s_j$ if $m_i$ is allocated to PM
$s_j$. We set $\mathbf{Y}(t) \triangleq
\Big[Y_1(t),\ldots,Y_{|V_m|}(t)\Big] \in \mathbb{Y}$ as the
aggregation of the locations of VMs, where $\mathbb{Y}$ is the set
of all possible $\mathbf Y(t)$. The
aggregation of the demands from all VMs can be represented as a vector
$\mathbf{R}(t) \triangleq \big[ R_1(t),\ldots,R_{|V_m|}(t)\big]$. We assume
that resource demand is quantized into $\Lambda$ discrete levels.
Hence, we define $R_l(t) \in \mathcal{R}$, where $\mathcal R =
\{r_0,r_1,\ldots,r_{\Lambda-1}\}$. 

We adopt the finite-state stationary Markov chain to model the resource
demand and to capture its temporal correlation. The model $\mathbf R(t)$ is
formulated with the following parameters,

              \begin{itemize}
                \item The state space of the resource demand is given by $\mathbf{R}(t) \in \mathbb{Q}$,
where $\mathbb{Q}=\{{\eta}_1,\ldots,{\eta}_{|\mathbb{Q}|}\}$. One item
$\mathbf{\eta}_i \in \mathbb{Q}$ is a vector of the demands from all VMs,
denoted as $\mathbf{\eta}_i = [\eta_{i,1},\ldots,\eta_{i,|V_m|}]$, where
$\eta_{i,l}$ is the demand from VM $m_l$ at state $\eta_i$.
                \item The transition kernel is given by
                \begin{align}\label{formula:approximate}
                    \phi_{i,j} &\triangleq Pr\Big[\mathbf{R}(t+1) = \eta_i | \mathbf{R}(t) = \eta_j\Big] \nonumber\\
                              & = \prod_{l=1}^{|V_m|} Pr\Big[R_l (t) = \eta_{i,l} | R_l(t-1) = \eta_{j,l}\Big].
                \end{align}
                Here we assume the demands from different VMs are independent.
                By defining
                \begin{equation}\label{formula:phiijl}
                \phi_{i,j,l} \triangleq Pr\Big[\eta_{i,l} | \eta_{j,l}\Big],
                \end{equation}
                \htan{we have} $\phi_{i,j} =\prod_{l=1}^{|V_m|} \phi_{i,j,l}$.
              \end{itemize}
            \begin{remark}[The motivation of the Markov chain model]
The resource demands of VMs may not follow a Markov chain strictly.
However, they would have some of the common features of Markov chain. For
example, the probability of the next CPU demand \hstan{partly} depends on
the request of the current slot, and this transition probability is
quasi-static within a coherent period. These features will be justified
in Section \ref{section:markov_model} by analyzing a real long-term
workload data trace from Google, which is frequently used in the
literature (e.g.
\cite{clusterdata:Zhang2014-Harmony,chen2014consolidating}). Therefore,
the Markov chain model of a VM's demands can be treated as an approximation
of the real world, capturing the temporal correlation of system dynamics.
\hfill\rule{6pt}{6pt}
            \end{remark}
       \subsection{Power Consumption Model}

The linear approximation model~\cite{buyya2010energy}, widely adopted in
the literature, is used to evaluate the power consumption of servers.
The power consumption of server $s_i$ is defined as
      \begin{equation}\nonumber
        P_{s_i}(t) = P_{idle} + (P_{max} - P_{idle})\cdot \min\{\frac{\sum_{\{l|Y_l(t) = s_i\}} R_l(t)}{T_r}, 1\},
      \end{equation}
where $P_{idle}$ is the power consumption when the server is in the idle
state (no computation task), and $P_{max}$ is the consumption in the
fully-loaded state (100\% utilization of CPU). If $s_i$ is not allocated
with any VM, it would be in sleep mode with a relatively low power
consumption, $P_{sleep}\ll P_{idle}$. The total power consumption of the
servers in the data center at time $t$ is
      \begin{equation}\label{formula:powerCluster}
        P_{total} (t) = \sum_{s_i \in V_s} P_{s_i}(t).
      \end{equation}
\vspace{-4mm}

Due to the dynamics of resource demands of VMs, the changes in power
consumption form a stochastic process. 
If some VMs have low resource demand with high probability, the
centralized manager can consolidate them into fewer PMs. In contrast,
when some VMs work at high resource demand with high probability, VM
migrations should be initiated to allocate more PMs to these VMs so that
resource shortage can be avoided.

      \subsection{VM migration}

VM migration, which is to move a running VM from~ one PM to another
without disconnecting the clients or applications, is a basic
operation supported by many platforms, such as Xen\cite{xen}
and KVM\cite{kvm}. In each time-slot, the data center manager
determines which VMs should be migrated and the PMs they should
migrate to.  We denote $ \gamma(t) \triangleq
\Big[\gamma_1(t),\ldots,\gamma_{|V_m|}(t)\Big] \in \mathcal{A}$ as the
migration of the VMs at the $t^{th}$ time-slot, where $\mathcal{A}$ is
the set of feasible migration actions and $\gamma_l(t)$ is the target PM
of $m_l$ after migration. We define $\mathcal A_l$ as the set of
available migration actions of $m_l$, i.e. $\gamma_l \in \mathcal A_l$.
We assume the migration can be finished in one time-slot. Thus, the
location of $m_i$ will be $\gamma_i$ in the next time-slot, i.e.
$Y_i(t+1) = \gamma_i$. If $m_i$ is not migrated at the $t^{th}$
time-slot, we have $Y_i(t) = Y_i(t+1) =\gamma_i(t)$.

VM migrations will result in many data transmissions consuming a large
amount of network bandwidth; so we set the maximum number of migrated VMs in one
time-slot as $T_m$, i.e.,
\begin{equation}\label{formula:restriction}
        \sum_{i=1}^{|V_m|} \mathbf{1}\Big[\gamma_i(t) \ne Y_i(t)\Big] \le¡¡T_m,
        \end{equation}
     where $\mathbf{1}\Big[\cdot\Big]$ is the indicator function which is equal to 1 when the condition holds, and 0 otherwise.

%

\subsection{Problem Formulation \label{section:problem}}
Our aim is to optimize the average power consumption and the resource
shortage over a long period given the dynamic resource demands from VMs.
Here, we formulate our VM management task
as an MDP problem, where the system state is defined as follows.

\begin{defn}[System State]{The system state at the $t^{th}$ time-slot can be uniquely specified by
            \begin{equation}\label{formula:cluster_state}
                \mathbf{S}(t) \triangleq \Big[ \mathbf{R}(t), \mathbf{Y}(t) \Big] \in \mathbb S,
             \end{equation} where $\mathbb{S} = \mathbb Q \times \mathbb Y$ denotes the space of the system states.
                }
        \end{defn}

At the beginning of each time-slot (say the $t^{th}$ time-slot), the data
center manager would collect the system information $\mathbf S(t)$ and
determine the control action for each VM.
            \begin{defn}[Stationary Control Policy]
              \label{defn:stationary}
A stationary control policy $\Omega$ is a mapping from the system state
$\mathbf{S}$ to the VM migration actions, i.e. $\Omega
\Big(\mathbf{S}(t)\Big) = \gamma(t)$ at the $t^{th}$ time-slot.
            \end{defn}
To quantify the resource shortage (i.e., the amount of resource demand that is not satisfied), we define the shortage level
$\theta_i$ for the $i^{th}$ PM at time $t$ as
      \begin{equation}\label{formula:shortage}
        \theta_i(t) \triangleq \max\{\frac{\sum_{\{ l | Y_l(t) = s_i\} } R_l(t)}{T_r} - 1,0\},
      \end{equation}

Instead of achieving instantaneous optimality, we focus on seeking
an optimal control policy to minimize the cost as a long-term average. The
total power consumption as a long-term average can be defined as follows,
     \vspace{-1mm} \begin{equation}\label{formula:powerExpected}
        \overline{P_{total}}(\Omega) \triangleq \lim_{T\rightarrow+\infty} \mathbb{E}_{\Omega}\Big[ \frac1T \sum_{t=1}^T P_{total}(t)\Big],
      \end{equation}
      where $\mathbb{E}_{\Omega}\Big[\cdot \Big]$ is the
      expectation under the stationary control policy $\Omega$.
Similarly, we denote $\overline\theta$ as the average (per-VM) resource
shortage level in long-term average, which is
    \vspace{-1mm}  \begin{equation}\label{formula:shortageExpected}
        \overline{\theta}(\Omega) \triangleq \lim_{T\rightarrow+\infty} \mathbb{E}_{\Omega}\Big[ \frac{1}{T\cdot |V_m|} \sum_{ i = 1 }^{|V_s|}\sum_{t=1}^T \theta_i(t)\Big].
      \end{equation}

Considering power consumption and resource shortage simultaneously, we
define the instantaneous cost by jointly combining the two objectives as
follows:
      \begin{equation}\label{formula:instantaneous_objective}
        g(t) \triangleq {P_{total}}(t) + \frac{\lambda}{|V_m|} \cdot \sum_{i=1}^{|V_s|}{\theta_i}(t),
      \end{equation} where $\lambda$ is the weight of resource
      shortage.\footnote{\hzhua{The weight} $\lambda$ indicates the relative importance of
      the resource shortage over the power consumption. It can be
interpreted as the corresponding Lagrange multiplier associated with
the resource shortage constraint. If the system does not allow resource
shortage, we can set $\lambda = +\infty$. We study the influence of the
setting of $\lambda$ in Section \ref{section:varyingLambda} by
simulations.}

      Finally, we define our problem as follows:
      \begin{problem}[The Dynamic VM Management Problem]\label{problem}
         \begin{align}
           &\min_{\Omega} ~~~~~ \overline{g}(\Omega) = \lim_{T\rightarrow +\infty} \mathbb E_{\Omega} \Big[\frac1T g(t)\Big] \label{problem:objective} \nonumber \\
           &          ~~~~~~~~~~~~~~~~ = \overline{P_{total}}(\Omega) + \lambda \cdot \overline{\theta}(\Omega)  \\
           &s.t.  ~~~~~~~(\ref{formula:restriction}) ~~~~~ \forall t
         \end{align}where the goal $\overline{g}(\Omega)$ is defined as
	 the long-term average cost of Eqn.
	 (\ref{formula:instantaneous_objective}) under the control policy
	 $\Omega$. The constraint is the maximum number of VM migrations
	 allowed in one time-slot.
      \end{problem}

For a given $\lambda$, the solution to Problem \ref{problem} corresponds
to a point in the Pareto optimal tradeoff curve between the average power
consumption 
and the average resource shortage. 

\section{\hzhua{MDP}-based Dynamic VM Management Algorithms\label{section:centralized-method}}
In this section, we propose an optimal MDP-based algorithm to solve the
dynamic VM management problem as just formulated. 
To avoid the curse of dimensionality, we derive another approximate
algorithm based on the optimal solution using the local dynamic resource
demand of each VM. We then prove the convergence of the proposed
algorithm and derive the bound of the approximation error compared with
the optimal.

      \subsection{Optimal VM Management Algorithm}


Since we focus on minimizing the average cost over time, we formulate the
dynamic VM management in Problem \ref{problem} as an infinite horizon
average MDP. The optimal solution can be achieved by solving the
equivalent Bellman's equation \cite{bertsekas1995dynamic}.

      \begin{lemma}[Equivalent Bellman's Equation] \label{lemma:bellman}
If a scalar $\beta$ and a vector of the utility function
      $\mathbf{V} = [V(S_1),V(S_2),\ldots]$ satisfy the Bellman's equation for Problem \ref{problem}, written as  $\forall S_i \in \mathbb S$,
        \begin{equation} \label{formula:bellman}
            \beta + V(S_i) = \min_{ \gamma \in \mathcal{A}(S_i)}\Big[ g(S_i) + \sum_{S_j \in \mathbb S}Pr[S_j|S_i,\gamma] V(S_j)
            \Big],
        \end{equation} where $g(S_i)$ is the instantaneous cost under the system state $S_i$,
        then $\beta$ is the optimal average cost:
        \begin{equation}\label{formula:bellmanCost}
          \beta = \min_{\Omega} ~\overline{g}(\Omega).
        \end{equation}
        Moreover, $\Omega$ is the optimal control policy if it attains the minimum in the R.H.S. of Eqn. (\ref{formula:bellman}).
        \hfill \rule{6pt}{6pt}
      \end{lemma}

Given the control policy $\Omega$, the transition probability of the system state $\mathbf S(t)$ at the $t^{th}$ time-slot can be written as
      \begin{align}\label{formula:Mchain_centra}
           &Pr\Big[\mathbf S(t+1) \Big| \mathbf S(t), \Omega(\mathbf S(t))\Big] \nonumber \\
          =&Pr\Big[\mathbf{R}(t+1),\mathbf{Y}(t+1)\Big| \mathbf{R}(t), \mathbf{Y}(t) , \Omega(\mathbf S(t))\Big]\nonumber\\
          =&Pr\Big[\mathbf{R}(t+1)\Big| \mathbf{R}(t)\Big]~Pr\Big[\mathbf{Y}(t+1) \Big| \mathbf{Y}(t), \Omega(\mathbf S(t))\Big]\nonumber\\
          =&Pr\Big[\mathbf{R}(t+1)=\eta_i\Big| \mathbf{R}(t)=\eta_j\Big] \nonumber\\
          =&\phi_{i,j} \quad \text{(the transition kernel defined in Eqn.(\ref{formula:approximate}))},
      \end{align}
where the third line means the migration action of VMs is deterministic,
i.e., $Pr\Big[\mathbf{Y}(t+1) | \mathbf{Y}(t), \Omega(\mathbf S(t))\Big] =
1$. Note that the transition kernel is unknown to the data center
manager. However, it can be learned by combining the sliding window
scheme and the maximum likelihood estimation (MLE)
\cite{beloglazov2013managing}. Recall that we assume the transition of
resource demand is independent among the VMs. We can capture the temporal
correlation of a VM's resource demands by estimating $\phi_{i,j,l}$ as defined
in Eqn. (\ref{formula:phiijl}). Denote the estimated $\phi_{i,j}$ and
$\phi_{i,j,l}$ as $\hat \phi_{i,j}$ and $\hat \phi_{i,j,l}$,
respectively. We can learn $\hat \phi_{i,j,l}$ as follows:
      \begin{equation} \label{formula:mle}
        \hat \phi_{i,j,l}(t) = \frac{\sum_{t'=t-T_w+1}^t \mathbf{1}[R_l(t'-1) = r_{i}]}{\sum_{t'=t-T_w+1}^t \mathbf{1}[R_l(t'-1) = r_{i} | R_l(t') = r_{j}]},
      \end{equation}
where $T_w$ is the length of the sliding window and $r_{i,l}$ is the demand level of $m_l$ in state $\eta_i$. 

Bellman's equation in Lemma \ref{lemma:bellman} is a fixed-point
problem in a functional space. Given the estimated transition
probability $\hat \phi_{i,j}$, the Bellman's equation can be solved
with value iteration or policy iteration \cite{Bellman:1957}, which
is a general solution to calculate the optimal utility function
iteratively. We demonstrate the procedure of value iteration to
compute the optimal utility function $\mathbf V$ in Algorithm
\ref{algo:VI}. In each iteration, the utility function will be
updated to $V^{\hat t}$ by finding the optimal policy under the
previous function $V^{\hat t -1}$ (Lines \ref{algo:updateStart} to
\ref{algo:updateEnd}). The function $\mathbf V$ will converge to
the optimal utility function which satisfies Lemma
\ref{lemma:bellman}. \htan{Before the convergence, it \drafty{has
been} proved that the utility function keeps increasing in each
iteration.} Thus, the utility function may converge to a large value. We
choose a reference state $S_r$ which can be any fixed state in the space
$\mathbb S$. In each iteration, the utility function for a state is
replaced by the relative value to that of the reference state (Line
\ref{algo:refStart} to Line \ref{algo:refEnd}). This operation can avoid
the utility function converging to a very large value, while not changing
the optimal control policy found after convergence. Due to the limited
space, readers can refer to \cite{bertsekas1995dynamic} for a better
understanding of value iteration. After obtaining the utility function,
we can find the optimal control policy by \hzhua{the R.H.S. of} Eqn. (\ref{formula:bellman}) in
Lemma \ref{lemma:bellman}.

      \begin{algorithm}[tbp]\label{algo:VI}
        \caption{Compute the Optimal Utility Function in Lemma \ref{lemma:bellman} by Value Iteration}
        \KwIn{The estimated transition probabilities $\hat\phi_{i,j}(t)$, $\forall i,j$ }
        \KwOut{The optimal utility function $V(\cdot)$}

        $\hat t = 0$, \quad      $V^0(S_i) = 0 , \forall S_i \in \mathbb{S}$\; \label{algo:initEnd}

        \While{Not converge}{\label{algo:convergeStart}
            $\hat t = \hat t+1$\;

            \For{$S_i \in \mathbb{S}$}{\label{algo:updateStart}
                $V^{\hat t}(S_i) = \min_{\gamma\in \mathcal{A}(S_i)} \Big\{g(S_i) + \sum_{S_j\in\mathbb{S}}Pr[S_j|S_i, \gamma]\cdot V^{\hat t-1}(S_j) \Big\}$\; \label{algo:updateEnd}
            }

             \For{$S_i \in \mathbb{S}$ and $S_i \ne S_r$}{\label{algo:refStart}
                $V^{\hat t}(S_i) = V^{\hat t}(S_i) - V^{\hat t}(S_r)$\;
            }
            $S_r = 0$\;\label{algo:refEnd}\label{algo:convergeEnd}
        }

        \end{algorithm}

It is NP-hard to solve the Bellman's equation \cite{cui2012survey}.
Algorithm \ref{algo:VI} traverses all the states in $\mathbb{S}$, and
needs exponential time and space to compute the utility function. We
propose an efficient approximate algorithm in the following subsections.

      \subsection{Per-VM Utility Function}
To avoid the curse of dimensionality, we have to reduce the state space
of the utility function. Instead of making all the states satisfy the
Bellman's equation in Eqn.~(\ref{formula:bellman}), we choose to satisfy
a few of the states, called the \emph{key states}. Before describing how to
find the key states, we define \emph{the feature state}, denoted as
$f_l(t)$, for VM $m_l$ at time $t$. The feature state $f_l(t)$ is the combination of
the location of $m_l$ and its expected resource demand in the steady
distribution of $R_l$ which is denoted as
      \begin{equation}\label{formula:steadyState}
          \pi_l^\infty = \lim_{n\rightarrow \infty} \mathbf \pi_l^0 (\mathbf P_l)^n,
        \end{equation}
where $\pi_l^0$ is a row vector ($\pi_l^0(i) = \hzhua{1}$ iff
$R_l = r_i$, and $\pi_l^0(i) = 0$ otherwise), and $\mathbf P_l$ is the
transition probability matrix of resource demand of $m_l$. Then, we have
      \begin{equation}\label{formula:feature_state}
      f_l(t) = \Big(\Big\lceil\sum_{i=0}^{\Lambda-1} r_i\cdot \pi_l^\infty (i)\Big\rceil, Y_l(t)\Big).
      \end{equation}

Based on the above, the key states for the VM $m_l$ at time-slot $t$,
denoted as $\mathbf S_K^l$, are defined as following,
      \begin{equation}\label{formula:keyStates}
        \mathbf S_K^l = \{i_{l,r,y}(t) | r = r_0,\ldots,r_{\Lambda-1} ; y
	= s_1,\ldots,s_{|V_s|} \},
      \end{equation} where $i_{l,r,y}(t) =
      \Big[S_1=f_1(t),S_2=f_2(t),\ldots,S_l=[r,y],\ldots,S_{|V_m|}=f_{|V_m|}(t)\Big]$
      denotes the state with $S_l = [r,y]$ and $S_i = f_i(t)~(\forall i
      \ne l)$. We denote $\mathbf S_K$ as the set of the key states
      satisfying the Bellman's equation in Eqn. (\ref{formula:bellman}),
      which is written as $\mathbf S_K = \bigcup_{l=1}^{|V_m|} \mathbf
      S_K^l$.

At each time-slot $t$, for the VM $m_l$, we define the per-VM utility
function $\widetilde{V}_l(i_{l,r,y}(t))$ as the utility function of the
key state $i_{l,r,y}(t)$, which satisfies, $\forall S_i \in \mathbf
S_K^l$
      \begin{equation}\label{formula:bellman_perVM}
        \beta_l + \widetilde{V_l}(S_i) = \min_{\gamma_l \in \mathcal A_l(S_i)} \Big\{ g(S_i) + \sum_{S_j \in \mathbf S_K^l} Pr[S_j | S_i, \gamma_l] \widetilde{V_l}(S_j) \Big\}.
      \end{equation}
      \htan{Here, $\beta_l$ is the optimal cost derived from the local
      key states $\mathbf S_K^l$ and the local control action $\mathcal
A_l$.} For ease of notation, we denote $\widetilde{V_l}(i_{l,r,y})$ as
$\widetilde{V_l}(r,y)$. We denote $\widetilde{\mathbf V}_l \triangleq
\Big[\widetilde{V}_l
(r_0,s_1),\ldots,\widetilde{V_l}(r_{\Lambda-1},s_{|V_s|})\Big]^T$ ($l =
1,\ldots,|V_m|$) as the per-VM utility function vector for all key states of each VM.

We adopt a feature-based method to approximate the utility function
$V(\mathbf S)$ as a linear summation of the per-VM utility functions, and
each VM updates its per-VM utility function with local state information
as well as the feature state of other VMs. Specifically, the proposed
linear approximation of the utility function $V(\mathbf S)$ is given by
      \begin{align}\label{formula:linearApproximation}
        V(\mathbf S) &= V([S_1,\ldots,S_{|V_m|}]) \nonumber\\
         & \approx \sum_{l=1}^{|V_m|}  \sum_{i = 0}^{\Lambda-1} \sum_{j = 1}^{|V_s|} \widetilde{V}_l([r_i,s_j]) \mathbf{1}[R_l =r_i, Y_l = s_j]\nonumber\\
         & = \mathbf{W}^T \mathbf F(\mathbf S),
      \end{align}
      where the parameter vector $\mathbf W$ and the feature vector $\mathbf F$ are given by Eqn. (\ref{formula:W}) and Eqn. (\ref{formula:F}), respectively.
      \begin{multline}\label{formula:W}
        \mathbf W \triangleq
	\Big[\widetilde{V}_0(r_0,s_1),\ldots,\widetilde
	V_0(r_{\Lambda-1},s_{|V_s|}),\\ \widetilde{V}_1(r_0,s_1)\ldots,\widetilde{V}_{|V_m|}(r_{\Lambda-1},s_{|V_s|})\Big]^T,
      \end{multline}\vspace{-8mm}
      \begin{multline}\label{formula:F}
        \mathbf F(\mathbf S) \triangleq \Big[~\mathbf{1} [R_0 = r_0, Y_0
		= s_1],\ldots,\\ \mathbf{1} [R_0 = r_{\Lambda-1}, Y_0 = s_{|V_s|}]
        ,\mathbf{1} [ R_{1} = r_{0}, Y_{1} = s_{1}], \\ \ldots ,\mathbf{1} [R_{|V_m|} = r_{\Lambda-1}, Y_{|V_m|} = s_{|V_s|}]~\Big]^T.
      \end{multline}

For example, suppose there are three VMs and two PMs. We set the
number of resource demand levels as $\Lambda=2$. At time-slot $t$, the
system state $\mathbf S(t)$ is $\Big[S_1(t) = [r_0,s_1], S_2(t) =
[r_1,s_1], S_3(t) = [r_0,s_2]\Big]$. Thus the utility function $V(\mathbf
S(t))$ can be given by the linear approximation, as follows:
      \begin{align}
        V(\mathbf S(t)) &\approx \widetilde{V}_1(S_1(t)) + \widetilde{V}_2(S_2(t)) + \widetilde{V}_3(S_3(t)) \nonumber\\
                             &= \widetilde{V}_1(r_0,s_1) + \widetilde{V}_2(r_1,s_1) + \widetilde{V}_3(r_0,s_2).\nonumber
      \end{align}
We can find that the state space is dramatically reduced from exponential
to polynomial with the linear approximation. However, the space of
control actions is still exponential due to the joint actions of all VMs.
In the following lemma, we exploit the equivalent Bellman's equation in
Eqn. (\ref{formula:bellman}) to show that the space of control actions
can also be reduced to polynomial by decomposing the joint control actions
into the individual control action of each VM.
      \begin{lemma} \label{lemma:apprixmate_action}
After the linear approximation in Eqn. (\ref{formula:linearApproximation}), the equivalent Bellman's equation in Eqn. (\ref{formula:bellman}) can be approximately computed as $\forall S_i \in \mathbb S$
      \begin{multline}
        \beta + V( S_i) = \min_{\gamma \in \mathcal A(\mathbf S_i)} \{g( S_i) + \sum_{S_j \in \mathbb S} Pr[S_j | S_i, \gamma] V(S_j)\}  \nonumber\\
          \approx \sum_{l=1}^{|V_m|}\Big( g(S_i) + \min_{\gamma_l \in \mathcal A_l(\mathbf S_i)} \{\sum_{S_j' \in \mathbf S^l_K} Pr[S_j' | S_i, \gamma_l] \widetilde V_l(S_j')\}\Big)
      \end{multline}
      \begin{proof}
Refer to Appendix A. 
      \end{proof}
      \end{lemma}

      \subsection{Approximate Algorithm: MadVM \label{section:madvm}}
Based on the above, we now describe our approximate
\underline{Ma}rkov-Decision-Process-based \underline{d}ynamic
\underline{VM} management algorithm for Problem~\ref{problem}, called
MadVM. We first assume there is a centralized manager that determines the
control policy, and then demonstrate how to implement our algorithm in
\hzhua{a distributed system}. MadVM consists of five main steps that are executed
sequentially, each in a time-slot:
      \begin{itemize}
         \item \textbf{Step 1, Initialization:} At the beginning of a
		 time-slot $t$, the centralized manager initializes the
		 per-VM utility function for each VM, i.e. $\widetilde
		 V_l(S_i) = 0$ ($l=1,\ldots,|V_m|, \forall S_i \in \mathbf S_K^l$). 
        \item \textbf{Step 2, Updating the Transition Probabilities:}
		Each VM updates the transition probabilities of the
		resource demand according to Eqn. (\ref{formula:mle}),
		and then determines its feature state $f_l$ according to
		Eqn. (\ref{formula:feature_state}).
        \item \textbf{Step 3, Information Collection of the Centralized
		Manager: } For each VM $m_l$, the manager collects its resource demand $R_l(t)$, the matrix of transition probability $\mathbf P_l$ and
		the feature state  $f_l(t)$.
        \item \textbf{Step 4, Calculating the Per-VM Utility Function:}
		\htan{Based on \drafty{the} feature states}, the
		centralized manager calculates the per-VM utility
		function $\mathbf{\widetilde V}_l$ for each VM $m_l$. The
		computation is similar to the optimal value iteration in
		Algorithm \ref{algo:VI} while the state space and utility
		function are replaced by the per-VM key states and per-VM
		utility function respectively, i.e. the operation from
		Line \ref{algo:updateStart} to Line \ref{algo:updateEnd}
		is replaced by  $\forall l = 1,\ldots,|V_m|,~\forall S_i \in \mathbf{S}_K^l$
            \begin{align}
              \widetilde V^{\hat t}(S_i) = &\min_{\gamma_l\in \mathcal{A}_l(S_i)} \Big\{g(S_i) + \\& \sum_{S_j\in\mathbf{S}_K^l}Pr[S_j|S_i, \gamma_l]\cdot  \widetilde V^{\hat t-1}(S_j) \Big\} \nonumber
            \end{align} And, the reference state $S_r$ in Line \ref{algo:refEnd} is replaced by state formed by the feature states of all VMs, i.e.
            \begin{equation}
              S_r = \Big[S_1=f_1(t),S_2=f_2(t),\ldots,S_{|V_m|}=f_{|V_m|}(t)\Big]\nonumber
            \end{equation}

        \item \textbf{Step 5, Determining the Control Actions:} We define the control utility $V_l^{c}$ for VM $m_l$ as the control action under the current system state $\mathbf S(t)$, which is
        \begin{align}\label{formula:biding_value}
          V_l^{c}(\mathbf S(t)) =& \min_{\gamma_l \in \mathcal A_l(\mathbf S(t))} \big\{ g(\mathbf S(t)) + \nonumber \\
                          &\sum_{S_j\in\mathbf S_K^l}Pr[S_j|\mathbf S(t), \gamma_l]\cdot \widetilde V_l(S_j) \big\},
        \end{align} The corresponding control action is the migration
	action which attains the minimum in the R.H.S. of Eqn.
	(\ref{formula:biding_value}).
         The centralized manager would rank the control utilities in
	 ascending order, and choose the top $T_m$ migration as the
	 control action in time $t$. Then, it goes back to Step 1 for
	 time $t+1$. \hfill\rule{6pt}{6pt}
      \end{itemize}

      We can see  MadVM has the following two distinct merits:
      \subsubsection{Low Complexity} 
In MadVM, only the local states and feature states are used to update
the per-VM utility function, and hence the state space is
$\Theta(|V_m||V_s|\cdot\Lambda)$. Moreover, the space of control actions
in Eqn. (\ref{formula:biding_value}) is also simplified from
$\Theta(|V_s|^{|V_m|})$ to $\Theta(|V_s|)$. Therefore, the complexity is
reduced significantly from exponential to polynomial.

\subsubsection{Implementation in \hzhua{ a distributed system}}  MadVM can be
implemented in a distributed system with a small amount of \hzhuaa{information sharing}. At the 3rd step of any time $t$, one VM $m_l$ shares
essential information to the other VMs, including its feature state
$f_l$ and the current state $S_l(t)$. Note that the matrix of transition probability  $\mathbf P_l$ need not be shared since the other VMs do not make use of it to update their utility functions. At the 4th step, each VM locally
calculates the per-VM utility function of itself. \htan{At the 5th step,
each VM, such as $m_l$, submits the control utilities $V_l^{c}(\mathbf
S(t))$ to hold an auction. The top $T_m$ VMs with the maximum control
utility win the auction and proceed with their control actions}.

      \subsection{Convergence Analysis}
Since MadVM only makes partial system states satisfy Bellman's equation
in Eqn. (\ref{formula:bellman}), the convergence of the approximate value
iteration is still unknown. In this section, we prove that the
approximate value iteration for updating per-VM utility function will
indeed converge.

For each VM $m_l$, let $\mathbf {\widetilde P}^l_\gamma$ denote the
transition probability matrix under the control action $\gamma$, i.e. the
$(i,j)$-th element of $\mathbf {\widetilde P}^l_\gamma$ is $Pr[S_i|S_j,
\gamma]$ ($\forall S_i, S_j \in \mathbf S_K^l$). We write $\mathbf
{\widetilde P}^l_\gamma$ as $\mathbf {\widetilde P}_\gamma$ without
ambiguity. We define the iteration operation (Line \ref{algo:updateEnd}
in Algorithm \ref{algo:VI}) as the mapping functions $\mathcal F$, and
$\mathcal F_{\gamma}$ is given as
      \begin{equation}\label{formula:optimalmappingFunction}
        \mathcal F (\widetilde {\mathbf V}_l) \triangleq \min_{\gamma } \{\mathbf{g} + \mathbf {\widetilde P}_\gamma \widetilde {\mathbf V}_l\}, \text{ and }
      \end{equation}\vspace{-0.4cm}
      \begin{equation}\label{formula:MappingFunction}
        \mathcal F_\gamma (\widetilde {\mathbf V}_l) \triangleq \mathbf g + \mathbf {\widetilde P}_\gamma \widetilde {\mathbf V}_l,
      \end{equation} where $\mathbf g =
      [g(i_{l,r_0,s_1}),\ldots,g(i_{l,r_{\Lambda-1},s_{|V_s|}})]$ is the
      vector of the cost of the local system states for VM $m_l$.
We denote the value of the system state $ S_i$ in $\mathcal
F_{\gamma_k}(\mathbf V)$ as $\mathcal F_{(k)}( S_i)$, where $\gamma_k$ is
the control action determined in the $k^{th}$ iteration. Therefore, the
value iteration can be given by
      \begin{equation}
        \widetilde{\mathbf V}_l^{k+1}( S_i) = \mathcal F_{(k)}( S_i) -  \mathcal F_{(k)}( S_r), ~~~\drafty{\forall S_i \in \mathbf S_K^l}.
      \end{equation}
We prove the convergence of the iterations \htan{to compute} the per-VM utility function with the following theorem. 

      \begin{theorem}[Convergence of the Per-VM Utility Function]
      \label{theorem:convergence}
      Let $\widetilde{\mathbf V}^*_l$ be the optimal average cost vector
      in the per-VM utility function of VM $m_l$. The sequence of utility
      function $\{\widetilde{\mathbf V}^k_l, k=1,2,\ldots\}$ in iterations converges to $\widetilde{\mathbf V}^*_l$ which satisfies
      \begin{equation}\label{formula:converge}
      \mathcal F(\widetilde{\mathbf V}_l^*(S_r))\mathbf e + \widetilde{\mathbf V}^*_l = \mathcal F(\widetilde{\mathbf V}^*_l),
      \end{equation} where $\mathbf e = (1,\ldots,1)^T$ is the unit vector.
      \begin{proof}
Refer to Appendix B. 
It is similar to the proof of the optimality of value iteration for the
undiscounted average cost problem in ~\cite{bertsekas1995dynamic}.
      \end{proof}
      \end{theorem}
   \vspace{-0.5cm}
    \begin{figure}[htbp]
      \centering
      \includegraphics[width=0.30\textwidth]{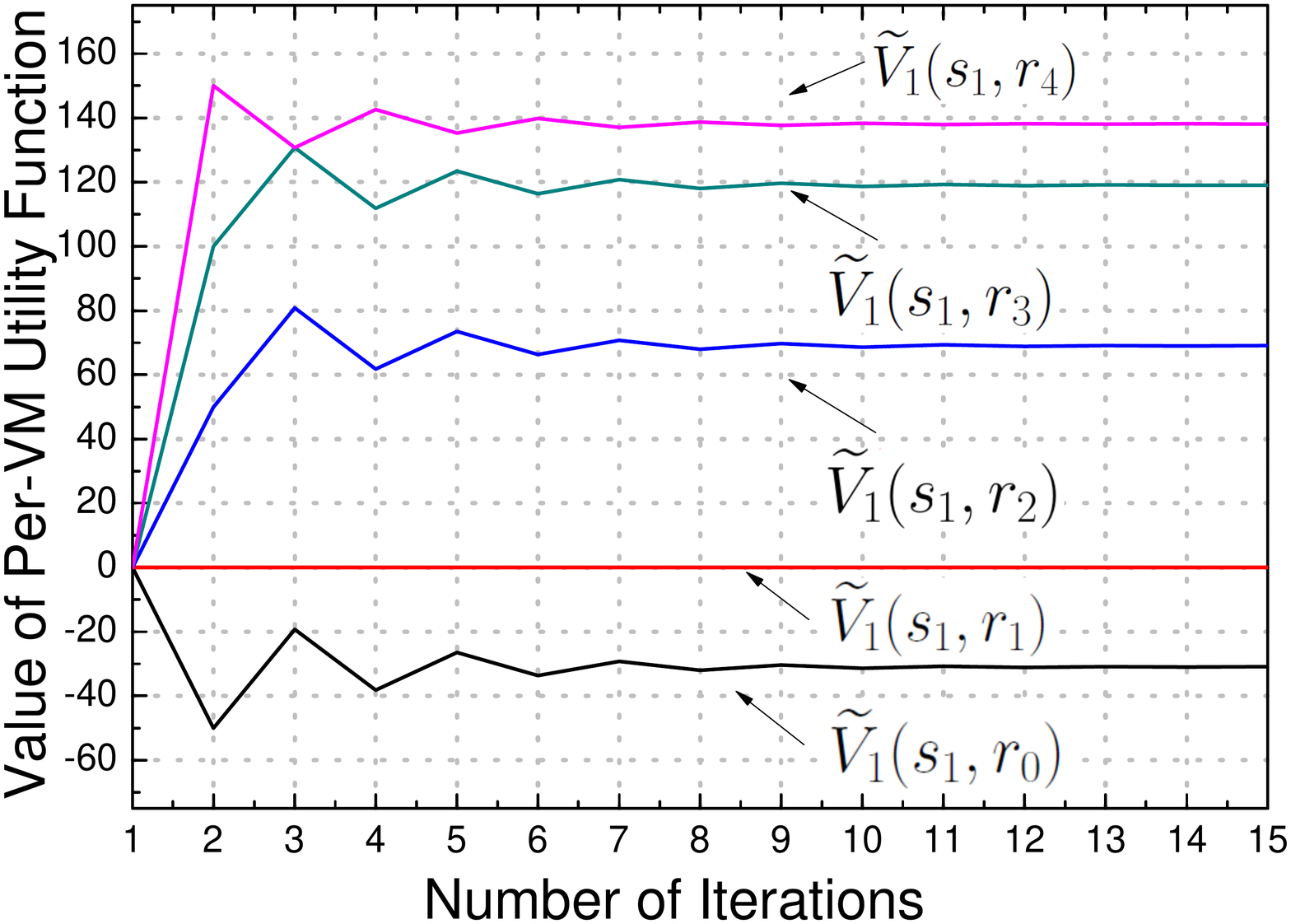}
      \vspace{-3mm}
      \caption{Illustration of convergence: Per-VM utility function
      versus the number of iterations. There is one PM and one VM, i.e. $
|V_s| = 1, |V_m| = 1$. The resource demand is quantized into 5 levels, i.e.
$\Lambda = 5$. }\label{figure:convergence}\vspace{-2mm}
    \end{figure}
Fig. \ref{figure:convergence} illustrates the convergence of the
per-VM utility function by an example.
We can see that the utility function converges before
the 15th iteration, and its value increases while the resource demand grows.

    \subsection{Asymptotic Performance Analysis}
We denote $\mathbf V^*$ as the vector of the optimal utility function for
each system state in the original system, which satisfies the equivalent
Bellman's equation in Lemma \ref{lemma:bellman}. Although Theorem
\ref{theorem:convergence} asserts the convergence of the per-VM utility
function, there is still an approximation error between the parameter
vector $\mathbf W$ and $\mathbf V^*$. Let $\mathbf M \in \mathbb R^{N_S
\times (|V_s|\cdot\Lambda)}$ denote the mapping matrix from $\mathbf W$
to the original system, where $N_S=(|V_s|\Lambda)^{|V_m|}$ is the number
of the original system states. We set $\mathbf V^\ddag$ as the
approximate utility function in the original system mapped from the
parameter vector $\mathbf W$. Here, for a matrix $\mathbf H$, we denote
its inverse as $\mathbf H^\dag$. Therefore, we have
      \begin{equation}\label{formula:relationVW}
        \mathbf V^\ddag = \mathbf M \mathbf W ~~~~\text{and}~~~~\mathbf W = \mathbf M^\dag \mathbf V^\ddag,
      \end{equation}
where $\mathbf M^\dag \in \mathbb R^{ |V_s|\Lambda \times N_S}$ has only
one 1 in each row and the positions of 1s correspond to the positions of
the key states.

In the following theorem, we provide a bound on the approximation error $|| \mathbf M \mathbf W - \mathbf V^*||$, where $||\cdot||$ is the $l^2$-norm of the vector.

      \begin{theorem}[Bound on the Approximation Error]
      \label{theorem:bound}
      Let $\mathbf X^*$ denote the optimal utility function after approximation which is $\mathbf X^* = arg\min_{\mathbf X} ||\mathbf M \mathbf X - \mathbf V^*|| = (\mathbf M^T\mathbf M)^{\dag}\mathbf M^T\mathbf V^*$. The approximation error is lower-bounded by
      \begin{equation}\label{formula:lowerbound}
        || \mathbf M \mathbf W - \mathbf V^*|| \ge || \mathbf M \mathbf X^* - \mathbf V^*||,
      \end{equation}
      and upper-bounded by
      \begin{equation}\label{formula:upperbound}
        || \mathbf M \mathbf W - \mathbf V^*|| \le \frac{a(c^n+1)}{1-\beta}|| \mathbf X^* - \mathbf{M}^\dag \mathbf V^*|| + || \mathbf M \mathbf X^* - \mathbf V^*||,
      \end{equation}
      where $a =  \sqrt{|V_m|\times (|V_s|\Lambda)^{|V_m|}}$, $n$ is integer and $0<\beta<1$. $n$ and $\beta$ should satisfy
      \begin{equation}\label{formula:uppercondition1}
        ||\mathcal F^n( \mathbf W) - \mathcal F^n( \mathbf X^*)||  \le \beta ||\mathbf W - \mathbf X^*||,
      \end{equation}
      and $c$ should satisfy
      \begin{multline}\label{formula:uppercondition2}
        ||\mathcal F^m( \mathbf W) - \mathbf M^\dag \mathbf V^*||   \\
        \le c ||\mathcal F^{m-1}( \mathbf W) - \mathbf M^\dag  \mathbf
	V^*||, m = 1,2,\ldots,n.
      \end{multline}
      \begin{proof}
       Refer to Appendix C. 
      \end{proof}
      \end{theorem}

Due to the convergence of the per-VM utility function in Theorem
\ref{theorem:convergence}, we have $\lim_{k\rightarrow\infty} \mathcal
F^k(\mathbf X^*) = \mathbf W$. Thus, there always exists a pair of
$(n,\beta)$ which satisfies $||\mathcal F^n(\mathbf W) - \mathcal
F^n(\mathbf X^*)|| \le \beta ||\mathbf W - \mathbf X^*||$. Intuitively,
the pair of $(n,\beta)$ measures the convergence speed of the value
iteration such that smaller $n$ or smaller $\beta$ results in higher
convergence speed.

Note that $||\mathcal F^m( \mathbf W) - \mathbf M^\dag \mathbf V^*|| =
||\mathbf M^\dag \mathcal F^\dag(\mathbf M \mathcal F^{m-1}(\mathbf X^*))
- \mathbf M^\dag \mathcal F^\dag(\mathbf V^*)||$, where $\mathcal F^\dag$
is a contraction mapping on the key states $\mathbf S_K$. There always
exists a sufficiently large $c \in [0,1)$ such that $||\mathcal F^m(
	\mathbf W) - \mathbf M^\dag \mathbf V^*|| \le c ||\mathcal
	F^{m-1}( \mathbf W) - \mathbf M^\dag  \mathbf V^*||$. The
	constant $c$ measures the contraction ratio of the contraction
	mapping $\mathcal F^\dag$, such that the smaller $c$ results in
	the larger contraction ratio.

In summary, if the value iteration operation $\mathcal F$ has good
convergence speed and the contraction mapping $\mathcal F^\dag$ has large
contraction ratio on the key states $\mathbf S_K$, we will have a small
upper-bound on approximation error. However, the approximation error can
never be smaller than $||\mathbf M \mathbf X^*-\mathbf V^*||$ owing to the
fundamental limitation on the vector dimension.

\section{Performance Evaluation \label{section:performance}}
In this section, we first analyze the temporal correlations of the
resource demands from VMs in Google's data centers. Then, we compare the
performance of our dynamic VM management algorithm MadVM with two
baseline algorithms using the data traces from Google Cluster
\cite{googlecluster} and PlanetLab~\cite{calheiros2011cloudsim}. The
former data trace has a long duration of 26 days, and the latter has more
intensely fluctuating resource demands from VMs.

\subsection{Resource Demand from VMs in Google Data Centers \label{section:markov_model}}
In MadVM, we adopt the Markov chain model to capture the temporal
correlation of VM resource demands approximately. Here, we
look into a real data trace \cite{googlecluster} coming from Google data
centers to validate the rationality of our model.
  \begin{figure}[htbp]
  \centering
\makeatletter\def\@captype{figure}\makeatother
\subfigure[Transition Probabilities among different levels ]{
\label{fig_heatmap}
\includegraphics[width=0.4\columnwidth]{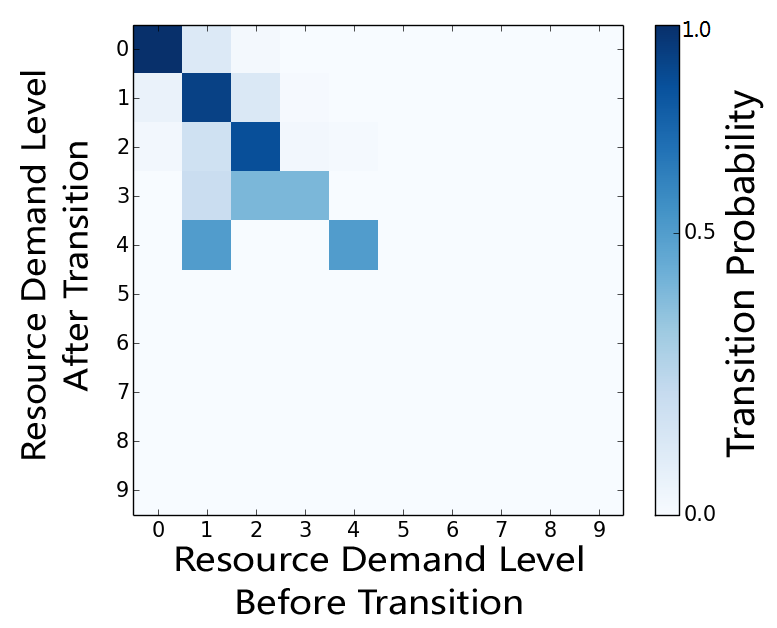}} \hspace{0.1 in}
\subfigure[$P_{2,2}$]{
\label{fig_transP22}
\includegraphics[width=0.48\columnwidth]{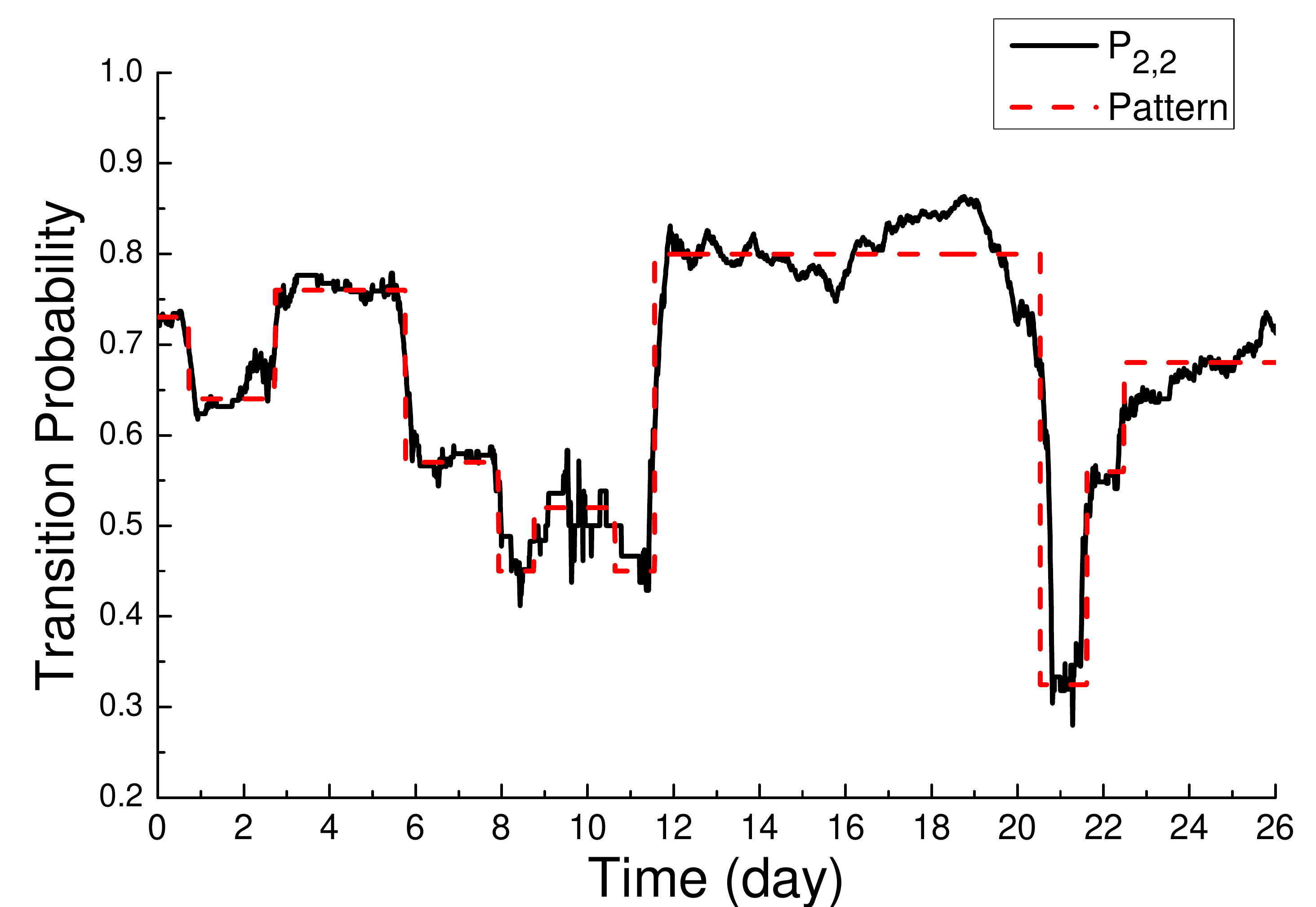}} \hspace{0.1 in}
\vspace{-4mm}
\caption{Transition probability of CPU demand in Google data trace}\label{fig_transition}
\vspace{-5mm}
\end{figure}

We arbitrarily choose one among 3000 VMs for observation. The experiment
results of other VMs are similar. Fig. \ref{fig_heatmap} shows the
transition probability matrix of CPU demand from the VM at a randomly selected
time-slot. We divide the CPU resource demand into 10 levels. The
transition probability is calculated by the maximum likelihood estimation
shown in the Eqn. (\ref{formula:mle}) with window size of 3 days. We
can find that the resource demand is always less than half of a PM's
capacity. Therefore, VM consolidation can evidently improve the
utilization of PMs. Moreover, \htan{the transition probabilities of the
demand level after transition are not uniformly distributed.} \hzhua{Therefore}, the
demand at time-slot $(t+1)$ is closely dependent on that at time-slot $t$. The
existence of temporal correlation proves the rationality of using the Markov
chain model to characterize the resource demand.

Fig. \ref{fig_transP22} shows the variation of $P_{2,2}$, the probability that the demand level stays on $r_2$ at two consecutive time-slots, which is the
the $3^{rd}$ row and the $3^{rd}$ column in Fig.
\ref{fig_heatmap}. Due to the limited space, we only take $P_{2,2}$ as an example here. Since the transition probabilities are
strongly correlated to the behavior of the VM,  the other transition
probabilities also have the same characteristics as $P_{2,2}$. We can
find that the transition probability always lingers around a value for
several days (drawn as the red dash line in Fig. \ref{fig_transP22}).
This phenomenon shows that the transition probability is quasi-static for
a short-term, e.g. 3 days. Note that the value iteration in our
algorithm MadVM will be executed every time-slot. \htan{Therefore, the
control policy capturing the temporal correlation can always be obtained
by our method according to the transition probabilities.}

\subsection{Performance Evaluation of MadVM}
We have conducted simulations to evaluate the performance of
MadVM using the VM utilization traces from Google Cluster and PlanetLab.
We compare MadVM to two baseline methods for dynamic VM management, a well-known algorithm CloudScale\cite{shen2011cloudscale} and a latest algorithm CompVM \cite{chen2014consolidating}.
Both CompVM and CloudScale predict future resource
demand to decide the control policy. In the initial VM allocation, MadVM
and CloudScale place each VM to the first-fit PM based on the expected VM
resource demand, while CompVM first predicts the pattern of resource
demand and then decide how to deploy the VMs to fully utilize the PMs.

In the experiments, one time-slot is set to 10 minutes. We configure the
PMs in the system with the capacity of a quad-core 1.5GHz CPU, and VMs with
the capacity of a single-core 1.5GHz. We set the powers for the fully-loaded
state $P_{max}$, the idle state $P_{idle}$ and the sleep state
$P_{sleep}$ to 500 watts, 250 watts and 50 watts, respectively. The
window size is set to 3 days to calculate the transition probabilities in
MadVM. The maximum number of VM migrations in one time-slot is set to 2\%
of the total number of VMs. We take the data of the first 6 days from the
two traces in the following experiments. The numbers of VMs and physical
machines are set as $1000$ and $500$ respectively. To evaluate the performance of the three methods, we use the following metrics
measured in 6 days and take the average over the time-slots spanned
(864 time-slots in total) : 1) \emph{the average power consumption:} to evaluate the energy efficiency; 2)\emph{the number of PMs used:} to illustrate the utilization of PMs; 3) \emph{the average resource shortage:} to illustrate how much resource demand is not satisfied; and 4) \emph{the average number of VM migrations:} to present the frequency of migrations.

In the following, we first demonstrate the impact of the parameter
$\lambda$, and then compare MadVM with CompVM and CloudScale  according to the above four metrics.

\subsection{Performance with different $\lambda$ \label{section:varyingLambda}}
\begin{figure}[htbp]\vspace{-4mm}
  \centering
\makeatletter\def\@captype{figure}\makeatother
\subfigure[Power Consumption]{
\label{fig_lambda_power}
\includegraphics[width=0.45\columnwidth]{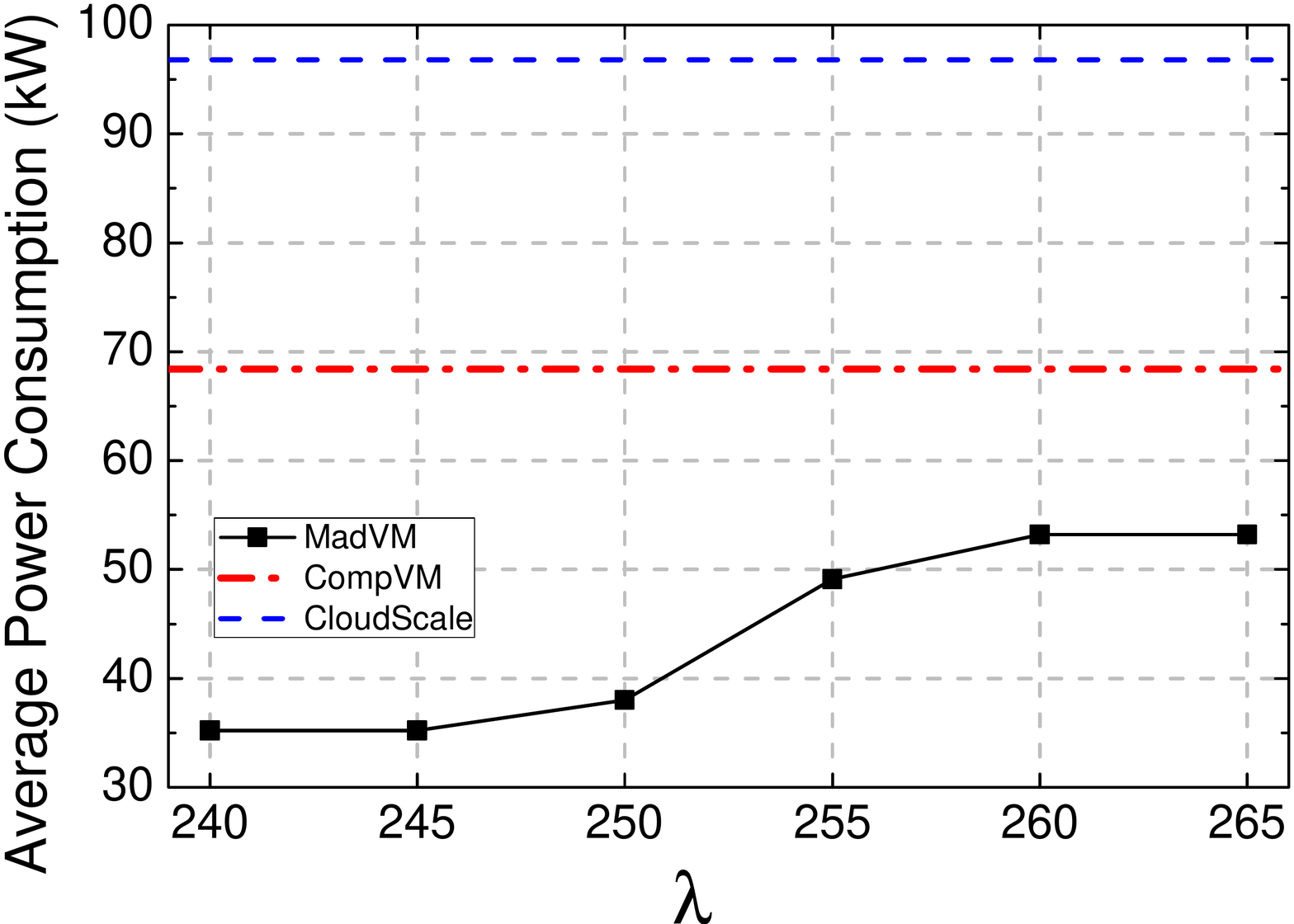}} \hspace{0.1 in}
\subfigure[Resource Shortage]{
\label{fig_lambda_shortage}
\includegraphics[width=0.45\columnwidth]{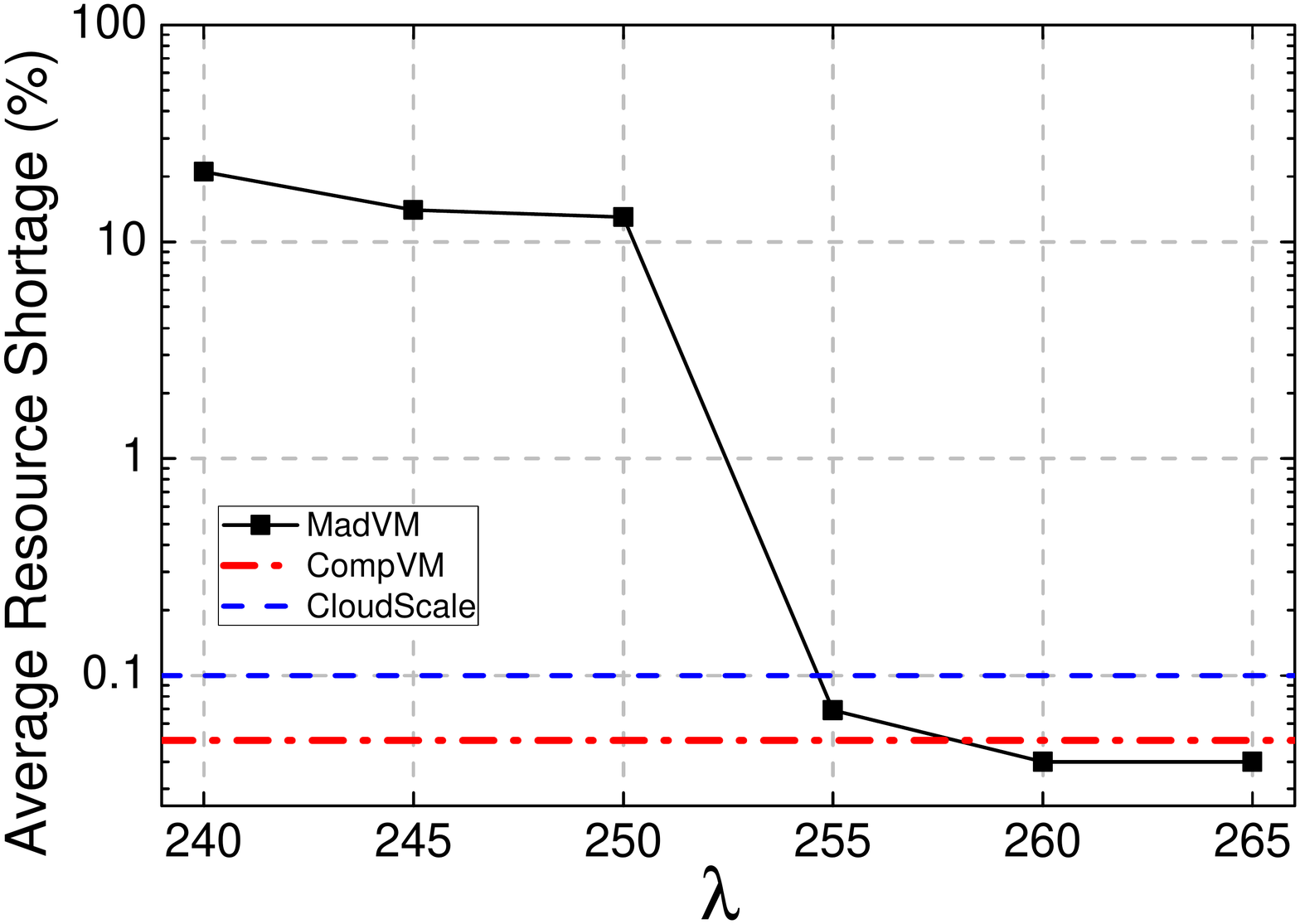}} \hspace{0.1 in}
\vspace{-4mm}
\caption{Power consumption and resource shortage with different $\lambda$ }\label{fig_lambda}
\vspace{-1mm}
\end{figure}
In order to handle the tradeoff between power consumption and resource
shortage, we introduce a parameter $\lambda$ in our objective defined in
Eqn. (\ref{formula:instantaneous_objective}).
Here, we investigate the relationship among $\lambda$, the average power
consumption and the average resource shortage with the Google cluster
trace. Fig.\ref{fig_lambda} indicates when $\lambda$ increases, the power
consumption increases and the resource shortage decreases, which is
reasonable since the larger $\lambda$, the greater the influence from the
shortage. We also include the power consumption of CompVM and CloudScale
under this data trace in Fig.\ref{fig_lambda_power}. We can see that
MadVM has the best power efficiency, even when $\lambda$ is set to a
large number. Fig.\ref{fig_lambda_shortage} demonstrates the resource
shortage when $\lambda$ changes (note that the vertical axis is in the
logarithmic scale). We can see that the resource shortage of MadVM decreases
dramatically when $\lambda$ increases. As the resource demands from VMs
are dynamic, resource shortage cannot be completely avoided. As shown
in Fig.\ref{fig_lambda_shortage}, CompVM and CloudScale also have a
shortage rate that is nearly 0. When $\lambda$ is large enough, i.e.,
$\lambda>260$, our shortage is the smallest.

In the following simulations, we set $\lambda$ to a very large value
(i.e., $\lambda=10^6$) so that our resource shortage is below the other
two baseline algorithms.

\subsection{Performance with different VM Resource Demand \label{section:varyingLoad}}
\begin{figure}[t]
\centering
\makeatletter\def\@captype{figure}\makeatother
\subfigure[Average power consumption]{
\label{fig_load_google_power}
\includegraphics[width=0.2\textwidth]{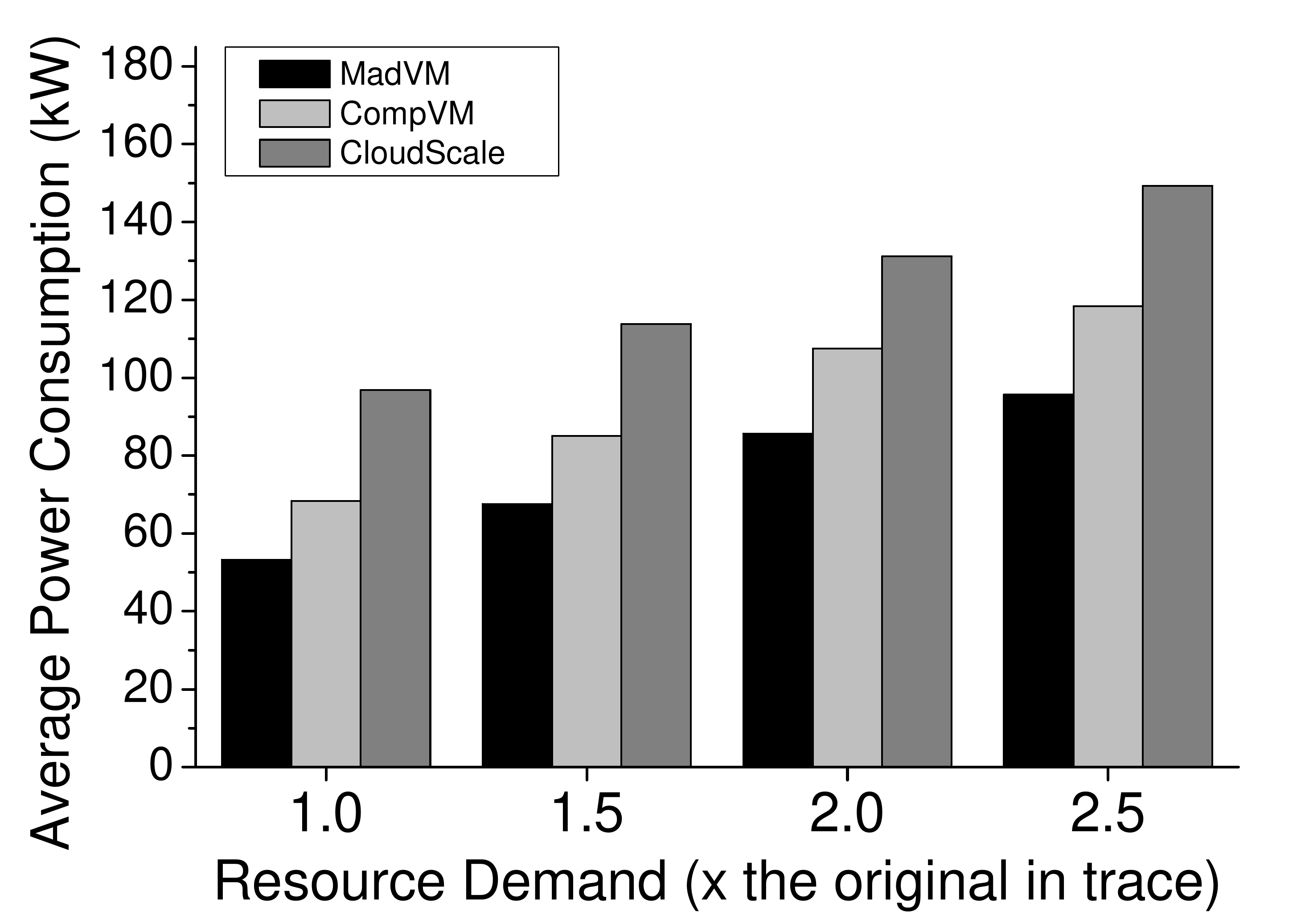}} \hspace{0.1 in}
\subfigure[\hzhua{Average} number of PMs used]{
\label{fig_load_google_pm}
\includegraphics[width=0.2\textwidth]{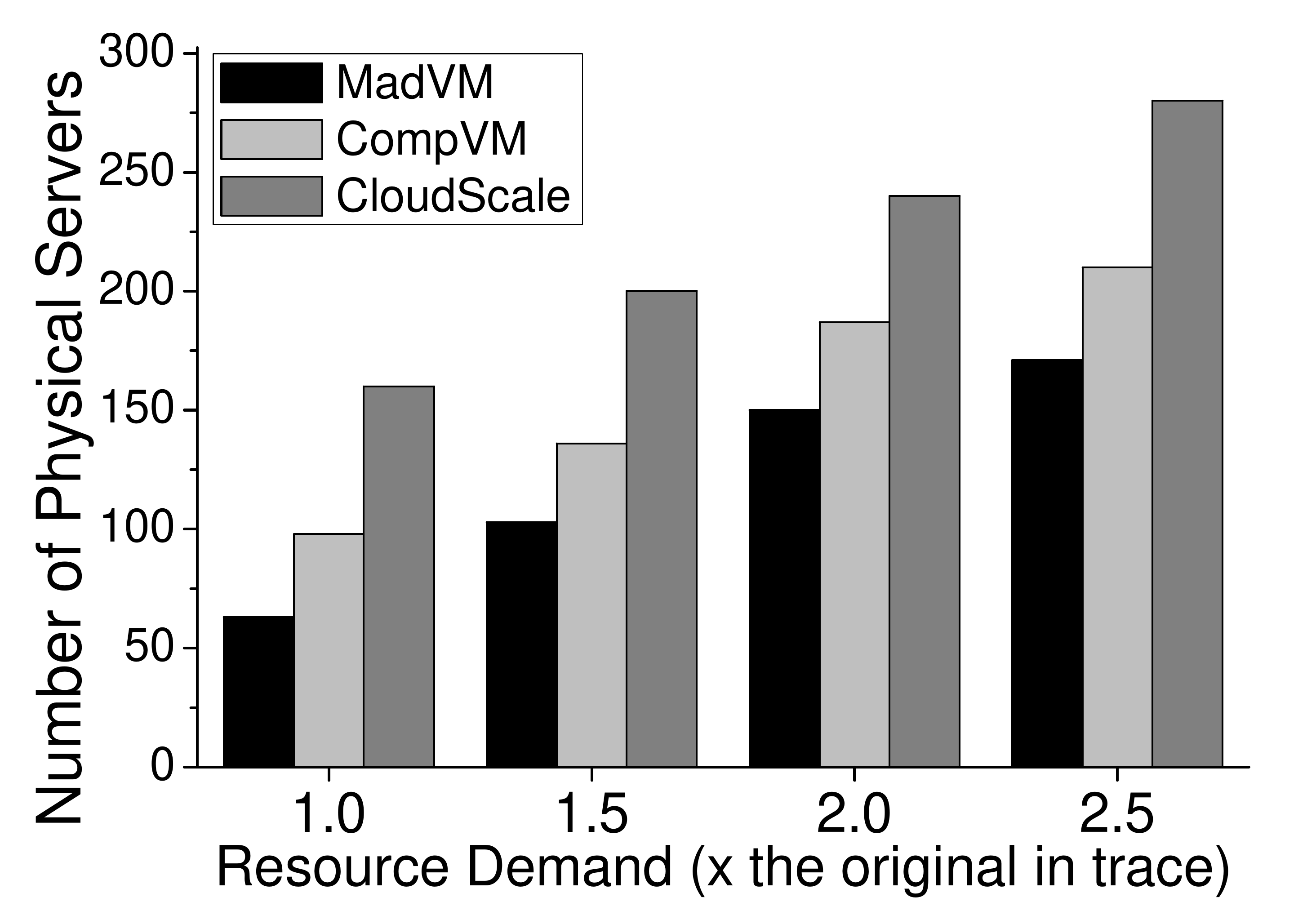}} \hspace{0.1 in}
\subfigure[Average resource shortage]{
\label{fig_load_google_shortage}
\includegraphics[width=0.2\textwidth]{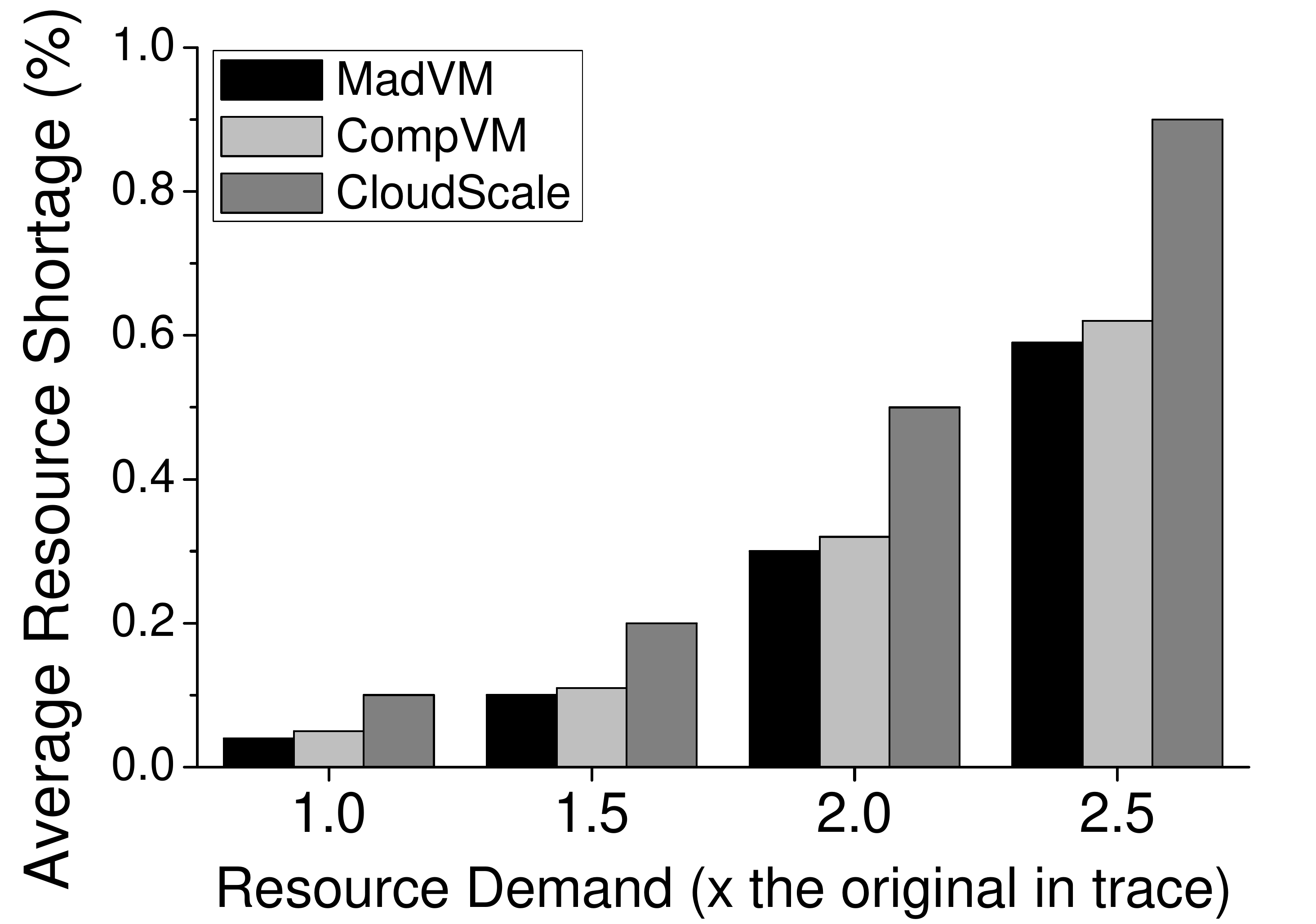}} \hspace{0.1 in}
\subfigure[Average VM migrations]{
\label{fig_load_google_migration}
\includegraphics[width=0.2\textwidth]{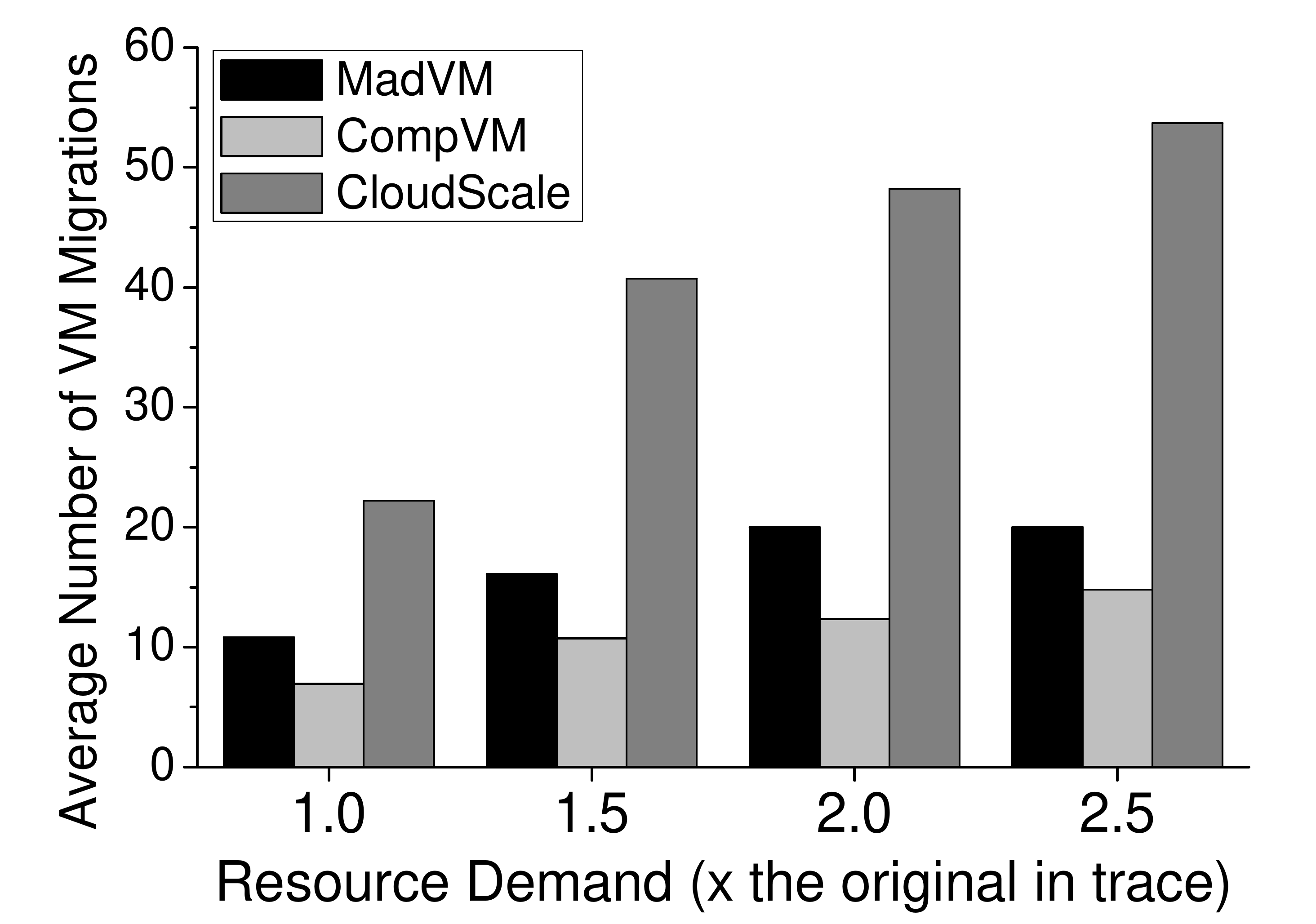}}
\vspace{-2mm}
\caption{\hzhua{Performance in the Google Cluster Trace}}
\vspace{-5mm}
\label{fig_google_load} \end{figure}

\begin{figure}[htbp]
\centering
\makeatletter\def\@captype{figure}\makeatother
\subfigure[Average resource shortage]{
\label{fig_load_plab_shortage}
\includegraphics[width=0.2\textwidth]{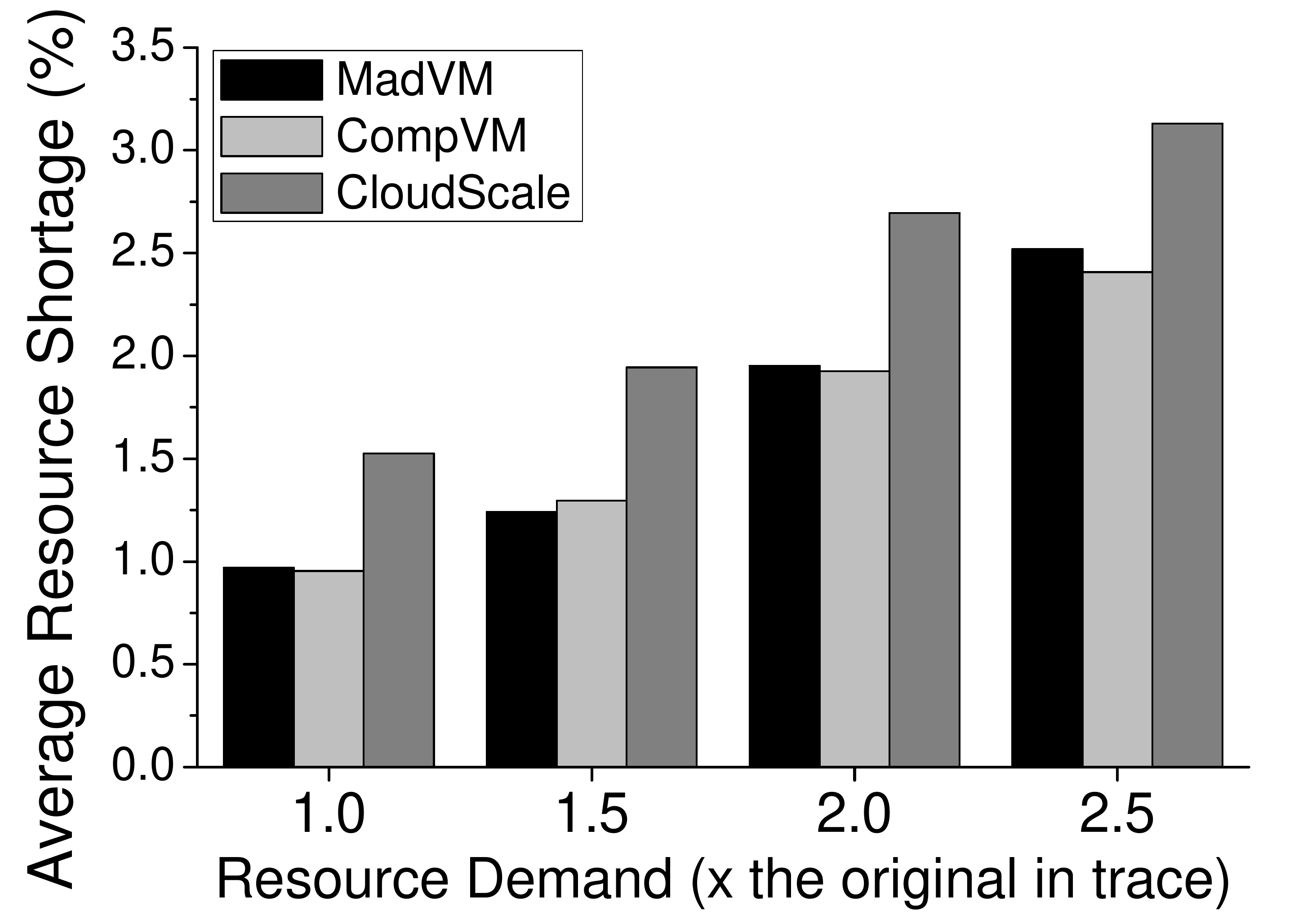}} \hspace{0.1 in}
\subfigure[Average VM migrations]{
\label{fig_load_plab_migration}
\includegraphics[width=0.2\textwidth]{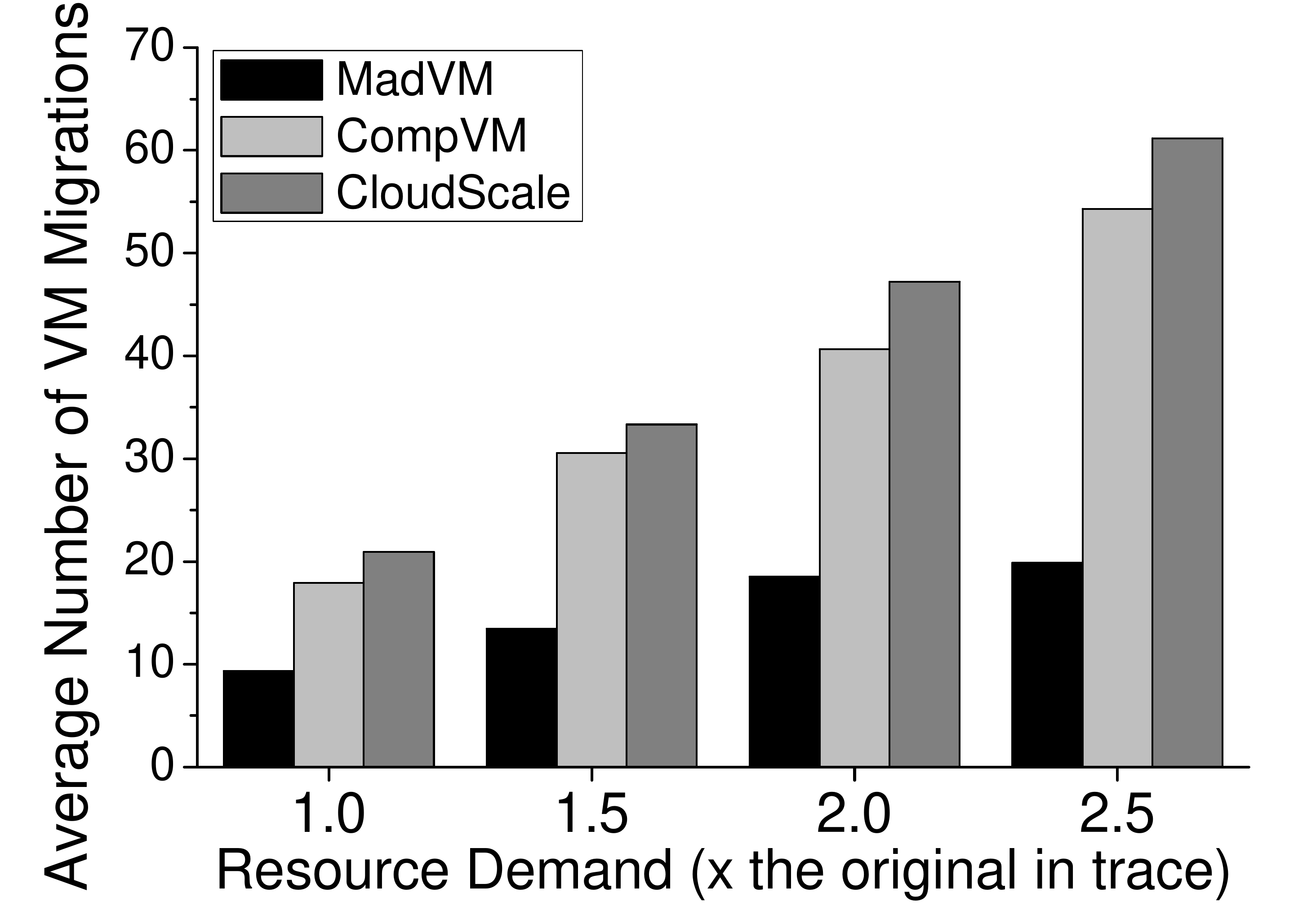}}
\vspace{-2mm}
\caption{\htan{Performance in the PlanetLab Trace}}
\vspace{-2mm}
\label{fig_plab_load} \end{figure}

Fig. \ref{fig_google_load} and \ref{fig_plab_load} demonstrate the
performance of three methods under different VM resource demands with
the Google trace and the PlanetLab trace. The resource demand of the VMs
is set to 1, 1.5, 2 and 2.5 times of the original in the traces.

Fig. \ref{fig_load_google_power} and Fig. \ref{fig_load_google_pm}
demonstrate the average power consumption and the average number of PMs
used per time-slot in the Google cluster trace respectively. In \hzhua{Fig.}
\ref{fig_load_google_power}, when the demand grows, the power consumption
increases in all three methods. MadVM has less power consumption than the
baseline methods, i.e. MadVM $<$ CompVM $<$ CloudScale. As
CompVM and CloudScale mainly consider the initial deployment with a
predicted pattern, they may result in failing to capture the dynamics of the
demand when the transition pattern changes. MadVM adopts an online learning
based approach to compute the transition probabilities, and the dynamics
can always be captured. In Fig. \ref{fig_load_google_pm}, while
increasing the demand, all methods use more PMs and hence consume more
power. MadVM uses the least PMs among the three, which also gives MadVM
$<$ CompVM $<$ CloudScale. This reflects that the resource utilization in
MadVM is the highest. The average power consumption and the number of PMs
used in the PlanetLab trace for the three methods are similar to the
results with the Google trace, i.e. MadVM $<$ CompVM $<$
CloudScale. We omit them here to save space.

\vspace{-2mm}

Fig. \ref{fig_load_google_shortage} and \ref{fig_load_plab_shortage} show
the average resource shortage over time in the two traces. While the
resource demand increases, the resource shortage grows in all three
methods (MadVM with a large $\lambda$, CompVM and CloudScale). They all
have a very low average resource shortage ($< 1\%$ \hzhuaa{in the Google trace and $< 3.5\%$ in the PlanetLab trace}) following the pattern MadVM
$\approx$ CompVM $<$ CloudScale. Because the resource demand in the
PlanetLab trace fluctuates more intensely than that in the Google trace, it is
harder to catch the dynamics. Hence all three methods show higher average
resource shortage in the PlanetLab trace.

Fig. \ref{fig_load_google_migration} and \ref{fig_load_plab_migration}
illustrate the average VM migrations per time-slot. As the
demand of VMs increases, the number of VM migrations also increases,
which follows  CompVM $<$ MadVM $<$ CloudScale in Fig.
\ref{fig_load_google_migration}. As both CompVM and CloudScale trigger VM
migrations when resource shortage appears, they migrate more VMs with the
increase of resource shortage. Instead of taking VM migration as a
remedy, MadVM takes it as a management method to utilize the resources
more efficiently. Thus, it is reasonable for MadVM to adopt more VM
migrations than CompVM \hzhuaa{in Fig.\ref{fig_load_google_migration}}. As demonstrated in
Fig.\ref{fig_load_plab_shortage}, the average resource shortage in
PlanetLab is relatively high, so the two baseline algorithms trigger many
more VM migrations. Thus, in Fig. \ref{fig_load_plab_migration}, the
result follows MadVM $<$ CompVM $<$ CloudScale. Therefore, we can say
MadVM manages the VMs more efficiently so that it needs fewer VM
migrations to keep to the same average resource shortage when the demand
fluctuates widely.

\ttan{In summary, while maintaining the resource shortage nearly to $0$, MadVM can reduce the power consumption by up to 23\% and 47\%  (averagely 19\% and 42\%) compared with CompVM and CloudScale, respectively. Moreover, the more widely the demands fluctuate, the better performance MadVM achieves in the power consumption and the frequency of VM migrations.}

\subsection{Resource Shortage under Insufficient Number of PMs \label{section:varyingPM}}
\begin{figure}[htbp]
\vspace{-3mm}
\centering
\makeatletter\def\@captype{figure}\makeatother
\subfigure[Google Cluster Trace]{
\label{fig_pm_google}
\includegraphics[width=0.44\columnwidth]{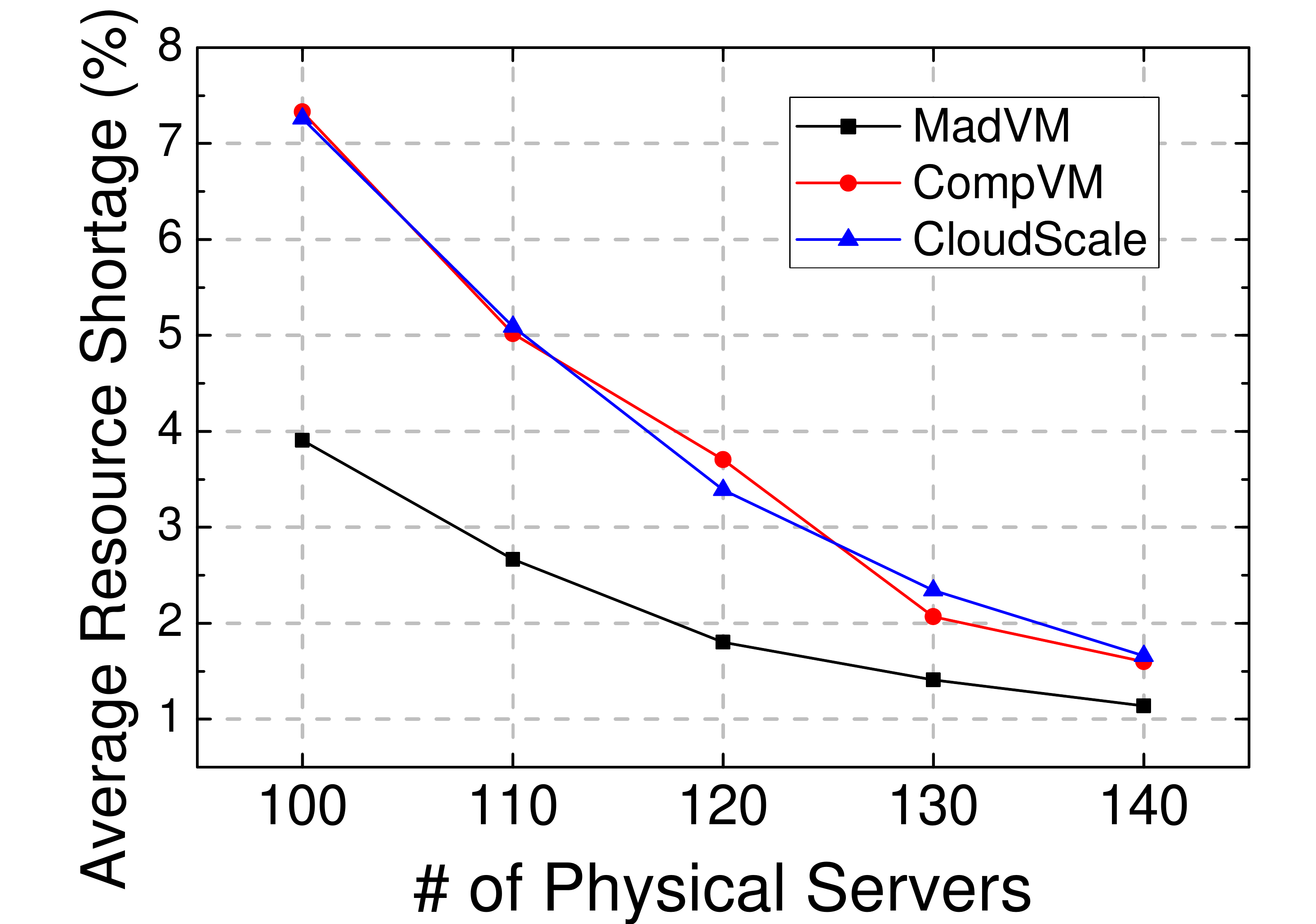}} 
\subfigure[PlanetLab Trace]{
\label{fig_pm_plab}
\includegraphics[width=0.44\columnwidth]{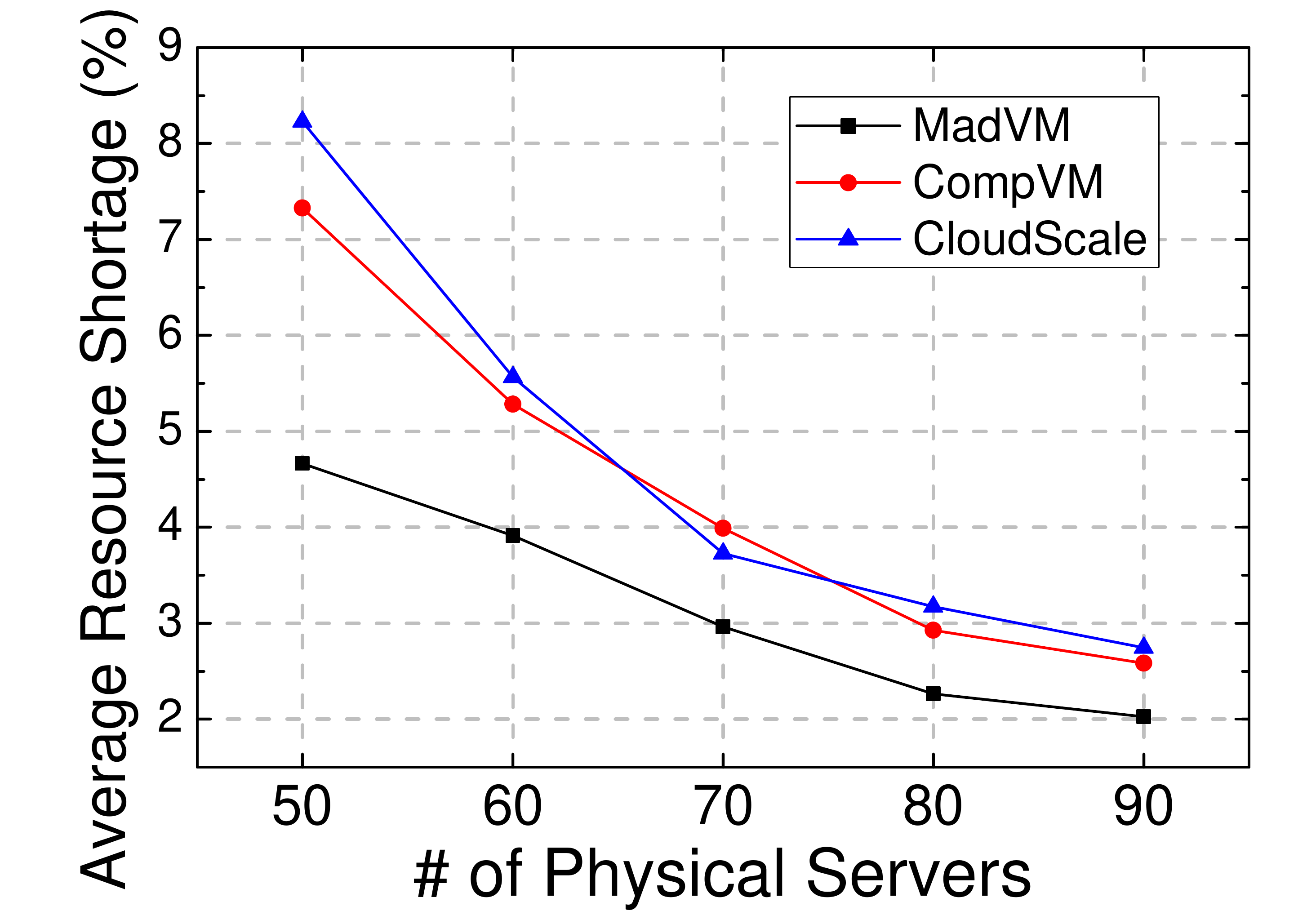}} 
\vspace{-3mm}
\caption{Average resource shortage under different number of PMs }
\vspace{-2mm}
\label{fig_shortage_lambda} \end{figure}
When setting a large $\lambda$ in MadVM, all three methods \hzhuaa{have very low} 
resource shortage. Here, we assume the number of PMs less than the demand of
the VMs, so that we can observe the resource shortage among the three
methods.
Figure \ref{fig_pm_google} and \ref{fig_pm_plab} show the resource
shortage under twice of the original resource demands in the traces. As
the number of PMs increases, the average resource shortage decreases as
expected. CompVM and ClouldScale have a similar resource shortage as they
adopt a similar greedy strategy when resource shortage appears. As
MadVM can capture the temporal correlation of the resource demand, the
resource shortage can be predicted and avoided. Thus, MadVM has a
significantly lower shortage, i.e., averagely 34\% smaller than the two baseline algorithms. Moreover, since MadVM adopts optimization
based approach to manage the VMs, it can find the optimal policy under local observation of each VM. Therefore, we can say MadVM has the most efficient utilization of the resources.

\section{Conclusion\label{section:conclusion}}
In this paper, we study dynamic VM management as a large-scale MDP
problem. By analyzing the Google workload trace, we show that the
resource demands of VMs are quasi-static over time. We first
derive an exponential-time optimal algorithm to solve the problem. Then,
we adopt approximate MDP and propose an efficient learning-based
approximate management approach, called MadVM. MadVM can capture the
temporal correlation of a VM's resource demand. Moreover, by
allowing a small amount of information sharing, MadVM can be implemented in
distributed fashion, which can help improve the robustness and
scalability of data centers.
The simulations based on two real-world workload traces show
significant performance gains by the proposed MadVM over two existing baseline
approaches in power consumption, resource shortage and the number of VM migrations.
Here, we consider only optimization of VM allocation. In the
future, we may consider a joint multi-level optimization including
CPU scheduling, the allocation of VMs and the cost of VM migration to
achieve energy saving in data centers.

\appendices
\section{Proof of Lemma \ref{lemma:apprixmate_action} \label{appendix:approximte_action}}
\begin{proof}
We define $S_{i,l}$ as the state of the VM $m_l$ in the state $S_i \in \mathbb S$.
By applying the linear approximation structure in Eqn. (\ref{formula:linearApproximation}), the equivalent Bellman's equation in Eqn. (\ref{formula:bellman}) can be written as follows, $\forall S_i \in \mathbb S$
      \begin{align*}
        \beta &+ V( S_i) = \min_{\gamma \in \mathcal A(\mathbf S_i)} \{g( S_i) + \sum_{S_j \in \mathbb S} Pr[S_j | S_i, \gamma] V(S_j)\}  \\
          &\approx \min_{\gamma \in \mathcal A(\mathbf S_i)} \{g( S_i) + \sum_{S_j \in \mathbb S} Pr[S_j | S_i, \gamma] \sum_{l=1}^{|V_m|} \widetilde V_l(S_{i,l})\}  \\
          &= g( S_i) + \min_{\gamma \in \mathcal A(\mathbf S_i)} \{ \sum_{l=1}^{|V_m|}\sum_{S_j \in \mathbb S} (\prod_{n=1}^{|V_m|}Pr[S_{j,n} | S_{i,n}, \gamma_n])  \widetilde V_l(S_{i,l})\}  \\
          &= g( S_i) + \min_{\gamma \in \mathcal A(\mathbf S_i)} \{ \sum_{l=1}^{|V_m|}\sum_{S_j \in \mathbb S} (Pr[S_{j,l} | S_{i,l},\gamma_l] \cdot \\ &~~~~~\prod_{n=1,n\ne l}^{|V_m|}Pr[S_{j,n} | S_{i,n}, \gamma_n])  \widetilde V_l(S_{i,l})\}\\
          &= g( S_i) + \min_{\gamma \in \mathcal A(\mathbf S_i)} \{ \sum_{l=1}^{|V_m|}\sum_{S_j' \in \mathcal S_K^l} (Pr[S_j' | S_{i},\gamma_l] \widetilde V_l(S_{j}')\}\\
          &= \sum_{l=1}^{|V_m|}\Big( g(S_i) + \min_{\gamma_l \in \mathcal A_l(\mathbf S_i)} \{\sum_{S_j' \in \mathbf S^l_K} Pr[S_j' | S_i, \gamma_l] \widetilde V_l(S_j')\}\Big).
      \end{align*} This completes the proof.
\end{proof}

\section{Proof of Theorem \ref{theorem:convergence} \label{appendix:proofConvergence}}
\begin{proof}
      For any admissible policy $\{\gamma_0, \gamma_1,\ldots\}$ there exists an $\epsilon > 0$ and a positive integer $m$ such that
      \begin{equation} \label{formula:proof_eps}
        [\mathbf {\widetilde P_{\mu_m}} \mathbf {\widetilde
	P_{\mu_{m-1}}}\ldots\mathbf {\widetilde P_{\mu_{1}}}]_{ir} \ge
	\epsilon ~~~~i = 1,\ldots,|\mathbb S|, \text{ and }
      \end{equation}
      \begin{equation} \label{formula:proof_eps2}
        [\mathbf {\widetilde P_{\mu_{m-1}}} \mathbf {\widetilde
	P_{\mu_{m-2}}}\ldots\mathbf {\widetilde P_{\mu_{0}}}]_{ir} \ge
	\epsilon ~~~~i = 1,\ldots,|\mathbb S|
      \end{equation} where $[\cdot]_{ir}$ denotes the element of $i^{th}$
      row and $j^{th}$ column of corresponding matrix. We denote
      $\delta^k(\mathbf S^i) = \widetilde{\mathbf V}^{k+1}_l(\mathbf S^i)
      - \widetilde{\mathbf V}^k_l(\mathbf S^i)$. We denote the $k^{th}$
      value iteration as follows:
      \begin{equation}
        \mathcal F_{(k)} (\mathbf S_i) = \min_{\gamma} \Big[g(\mathbf S_i) + \sum_{j} Pr[\mathbf S_j | \mathbf S_i , \gamma] \widetilde{\mathbf V}^k_l(\mathbf S_j) \Big].
      \end{equation} Set $\lambda_{(k)} = \mathcal F_{(k)}(\mathbf S_r)$.
      Then we have
      \begin{equation}
         \widetilde{\mathbf V}_l^{k+1} = \mathbf g + \mathbf{\widetilde P_{\gamma_k}} \widetilde{\mathbf V}_l^{k} - \lambda_k \mathbf e \le g + \mathbf{\widetilde P_{\gamma_{k-1}}} \widetilde{\mathbf V}_l^{k} - \lambda_k \mathbf e,
      \end{equation}
      \begin{equation}
         \widetilde{\mathbf V}_l^{k} =\mathbf g + \mathbf{\widetilde P_{\gamma_{k-1}}} \widetilde{\mathbf V}_l^{k-1} - \lambda_{k-1} \mathbf e \le g + \mathbf{\widetilde P_{\gamma_{k}}} \widetilde{\mathbf V}_l^{k-1} - \lambda_{k-1} \mathbf e.
      \end{equation} With the definition $\delta^k(\mathbf S^i) = \widetilde{\mathbf V}^{k+1}_l(\mathbf S^i) - \widetilde{\mathbf V}^k_l(\mathbf S^i)$, we obtain
      \begin{equation} \label{formula:proof_manyP}
        \mathbf{\widetilde P_{\gamma_{k}}} \delta^{k-1}+ (\lambda_{k-1} - \lambda_k)\mathbf e \le \delta^k \le
         \mathbf{\widetilde P_{\gamma_{k-1}}} \delta^{k-1}+ (\lambda_{k-1} - \lambda_k)\mathbf e.
      \end{equation} By iterating, we have
      \begin{multline}
        \mathbf{\widetilde P_{\gamma_{k}}\ldots\widetilde P_{\gamma_{k-m+1}}} \delta^{k-m}+ (\lambda_{k-m} - \lambda_k)\mathbf e \le \delta^k \\ \le
         \mathbf{\widetilde P_{\gamma_{k-1}}\ldots\widetilde P_{\gamma_{k-m}}} \delta^{k-m}+ (\lambda_{k-m} - \lambda_k)\mathbf e.
      \end{multline} Due to (\ref{formula:proof_eps}) and (\ref{formula:proof_eps2}), the R.H.S. of (\ref{formula:proof_manyP}) yields
      \begin{equation}
        \delta^k(\mathbf S_i) \le \sum_{\mathbf S_j\in \mathbb S}
	[\mathbf{\widetilde P_{\gamma_{k}}\ldots\widetilde P_{\gamma_{k-m+1}}}]_{ij} \delta^{k-m}(\mathbf S_i) + \lambda_{k-m} - \lambda_k
      \end{equation}
      \begin{equation}
        \Rightarrow \delta^k(\mathbf S_i) \le (1-\epsilon) \max_j \delta^{k-m}(\mathbf S_j) + \lambda_{k-m} - \lambda_k.
      \end{equation} Therefore, we obtain
      \begin{equation} \label{formula:proof_RHS}
        \max_j \delta^k(\mathbf S_j) \le (1-\epsilon) \max_j \delta^{k-m}(\mathbf S_j) + \lambda_{k-m} - \lambda_k.
      \end{equation} Similarly, from the L.H.S. of (\ref{formula:proof_manyP}) we obtain
      \begin{equation}\label{formula:proof_LHS}
        \min_j \delta^k(\mathbf S_j) \ge (1-\epsilon) \min_j \delta^{k-m}(\mathbf S_j) + \lambda_{k-m} - \lambda_k.
      \end{equation} By combining (\ref{formula:proof_RHS}) and (\ref{formula:proof_LHS}), we have
      \begin{multline}
        \max_i \delta^k (\mathbf S_i) - \min_i \delta^k(\mathbf S_i) \le \\ (1-\epsilon) (\max_i \delta^{k-m} (\mathbf S_i) - \min_i \delta^{k-m}(\mathbf S_i) )
      \end{multline} For some $B>0$ and all $k$, we have
      \begin{equation}
        \max_i \delta^k (\mathbf S_i) - \min_i \delta^k(\mathbf S_i) \le B(1-\epsilon)^{k/m}.
      \end{equation} Since $\delta^k(\mathbf S_i) = 0$, it follows that
      \begin{multline}
        |\mathbf{\widetilde V}^k_l(\mathbf S_i) - \mathbf{\widetilde V}^k_l(\mathbf S_i)| = |\delta^k(\mathbf S_i)| \le\\
        \max_j \delta^k(\mathbf S_j) - \min_j \delta^k(\mathbf S_j) \le B(1-\epsilon)^{k/m}.
      \end{multline} Therefore, for every $n>1$ and $\mathbf S_i$ we have
      \begin{align}
        |\mathbf{\widetilde V}^{k+n}_l(\mathbf S_i) - \mathbf{\widetilde V}^k_l(\mathbf S_i)| &\le \sum_{j=0}^{n-1}|\mathbf{\widetilde V}^{k+j+1}_l(\mathbf S_i) - \mathbf{\widetilde V}^{k+j}_l(\mathbf S_i)| \nonumber\\
        & \le B(1-\epsilon)^{k/m} \sum_{j=0}^{n-1} (1-\epsilon)^{j/m} \nonumber \\
        & = \frac{B(1-\epsilon)^{k/m}(1-(1-\epsilon)^{n/m})}{1-(1-\epsilon)^{1/m}},
      \end{align} so ${\mathbf{\widetilde V}^{k}_l(\mathbf S_i)}$ is a Cauchy sequence and converges to a limit $\mathbf{\widetilde V}^{*}_l(\mathbf S_i)$. Therefore, we obtain the equation (\ref{formula:converge}). This completes the proof.
      \end{proof}

\section{Proof of Theorem \ref{theorem:bound} \label{appendix:bound}}
The lower-bound is straightforward. The proof of the upper-bound is given
below. Since

\begin{align*}
  ||\mathbf W^\infty - \mathbf X^*|| &\le ||\widetilde{\mathbf F}^n(\mathbf W^\infty) -\widetilde{\mathbf F}^n (\mathbf X^*)||+||\widetilde{\mathbf F}^n( \mathbf X^*) - \mathbf X^*|| \\
                                     &\le \beta ||\mathbf W^\infty - \mathbf X^*|| + ||\widetilde{\mathbf F}^n(\mathbf X^*) - \mathbf X^*||,
\end{align*}
we have
\begin{equation}\label{formula:mergeBeta}
  ||\mathbf W^\infty - \mathbf X^*|| \le \frac{1}{1-\beta} ||\widetilde{\mathbf F}^n(\mathbf X^*) - \mathbf X^*||.
\end{equation}
From the definition of constant $c$, we have
\begin{align}\label{formula:mergeC}
  ||\widetilde{\mathbf F}^n(\mathbf X^*) - \mathbf X^*|| &\le ||\widetilde{\mathbf F}^n(\mathbf X^*) - \mathbf M^\dag \mathbf V^*|| + ||\mathbf M^\dag \mathbf V^* - \mathbf X^*|| \nonumber\\
                                                     &\le c||\widetilde{\mathbf F}^{n-1}(\mathbf X^*) - \mathbf M^\dag \mathbf V^*|| + ||\mathbf M^\dag \mathbf V^* - \mathbf X^*||\nonumber\\
                                                     &\le (c^n+1) ||\mathbf M^\dag \mathbf V^* - \mathbf X^*||.
\end{align}
As a result,
\begin{align*}
  ||\mathbf {MW}^\infty - \mathbf V^*|| & \le ||\mathbf {MW}^\infty - \mathbf {MX}^*|| + ||\mathbf{MX}^*-\mathbf V^*||\\
                                             & \le a||\mathbf W^\infty - \mathbf X^*|| +  ||\mathbf{MX}^*-\mathbf V^*||\\
                                             & \le \frac{a(c^n+1)}{1-\beta} ||\mathbf M^\dag \mathbf V^*- \mathbf X^*|| \\
                                             &~~~ + ||\mathbf {MX}^* - \mathbf V^*||,
\end{align*}
where the last inequality is because of (\ref{formula:mergeBeta}) and (\ref{formula:mergeC}). This completes the proof.

\bibliographystyle{IEEEtran}

\end{document}